\documentclass[a4paper,11pt]{article}

\pdfoutput=1 

\usepackage{jheppub} 

\usepackage{amsmath,amsbsy,amssymb}
\usepackage{graphicx}
\usepackage{dcolumn}
\usepackage{bm}
\usepackage{bbm}
\usepackage{epsfig}
\usepackage{slashed}
\usepackage{color}
\usepackage{fixmath}
\usepackage[font=small]{caption}
\usepackage[font=small]{subcaption}
\usepackage{slashed}
\newcommand{\genericT}{\ensuremath{T}}

\usepackage[T1]{fontenc} 

\usepackage{booktabs}

\makeatletter

\newcommand{\newc}{\newcommand}
\newc{\renewc}{\renewcommand}
\newcommand{\met}{\slashed{E}_\genericT}

\def\beq{\begin{equation}}
\def\eeq{\end{equation}}
\def\bea{\begin{eqnarray}}
\def\eea{\end{eqnarray}}
\def\bitem{\begin{itemize}}
\def\eitem{\end{itemize}}
\def\ba{\begin{array}}
\def\ea{\end{array}}
\def\bal{\begin{align}}
\def\eal{\end{align}}
\def\bi{\begin{itemize}}
\def\ei{\end{itemize}}
\def\lsim{\mathrel{\rlap{\lower4pt\hbox{\hskip1pt$\sim$}}
    \raise1pt\hbox{$<$}}}         
\def\gsim{\mathrel{\rlap{\lower4pt\hbox{\hskip1pt$\sim$}}
    \raise1pt\hbox{$>$}}}

\def\pb{\textrm{\:pb}}
\def\fb{\textrm{\:fb}}
\def\fbi{\textrm{\:fb}^{-1}}
\def\gev{\textrm{\:GeV}}
\def\tev{\textrm{\:TeV}}

\def\trm{\textrm}
\def\mrm{\mathrm}
\def\srm#1{{\trm{\tiny #1}}}
\def\mbx#1{{\mrm{\scalebox{0.6}{#1}}}}

\newcommand{\GOSAM}{{\textsc{Go\-Sam }}}

\newcommand{\ttbar}{$t\bar{t}$ }

\preprint{{\small DESY 14-177\\\vphantom{.}\hfill MPP-2014-371}}

\title{Model-Independent Production of a Top-Philic Resonance at the LHC}

\author[a,b]{Nicolas Greiner,}
\author[c]{Kyoungchul Kong,}
\author[c,d]{Jong-Chul Park,}
\author[d,e]{Seong Chan Park,}
\author[b]{Jan-Christopher Winter\,}

\affiliation[a]{DESY Theory Group, Notkestr.~85,
  D-22607 Hamburg, Germany}
\affiliation[b]{Max Planck Institut f\"ur Physik, F\"ohringer Ring 6,
  D-80805 M\"unchen, Germany}
\affiliation[c]{Department of Physics and Astronomy, University of
  Kansas, Lawrence, KS 66045, USA}
\affiliation[d]{Department of Physics, Sungkyunkwan University, Suwon
  440-746, Korea}
\affiliation[e]{School of Physics, Korea Institute for Advanced Study,
  Seoul 130-722, Korea}

\emailAdd{greiner@mpp.mpg.de}
\emailAdd{kckong@ku.edu}
\emailAdd{jcpark@ku.edu}
\emailAdd{s.park@skku.edu}
\emailAdd{jwinter@cern.ch}

\abstract{
We investigate the collider phenomenology of a color-singlet vector
resonance, which couples to the heaviest quarks, the top quarks,
but very weakly to the rest of the fermions in the Standard Model.
We find that the dominant production of such a resonance does not
appear at the tree level -- it rather occurs at the one-loop level in
association with an extra jet.
Signatures like $t\bar t$ plus jets readily emerge as a result of the
subsequent decay of the resonance into a pair of top quarks.
Without the additional jet, the resonance can still be produced
off-shell, which gives a sizeable contribution at low masses. The
lower top quark multiplicity of the loop induced resonance production
facilitates its reconstruction as compared to the tree
level production that gives rise to more exotic signatures involving
three or even four top quarks in the final state.
For all these cases, we discuss the constraints on the resonance
production stemming from recent experimental measurements in the top
quark sector. We find that the top-philic vector resonance remains
largely unconstrained for the majority of the parameter space,
although this will be scrutinized closely in the Run\:2 phase of the
LHC.}


\begin{document} 
\maketitle
\flushbottom

\section{Introduction}

With the recent discovery of a Higgs boson at the Large Hadron
Collider (LHC) at CERN, we are beginning to explore the physics of the
TeV scale in earnest.
The next goal is the precise measurement of the properties of the
newly discovered particle, and to find new physics beyond the Standard
Model.
The lightness of the discovered Higgs boson and precision measurements
of Standard Model (SM) physics
provide important guidance on building models beyond the SM.
So far, the LHC has not seen any indication of new physics phenomena and 
the null results at the first run of the LHC provide stringent
constraints on the mass scale of the new physics.
Naively, one could imagine a scenario where the particle spectrum is
relatively degenerate, or where new particles are simply out of reach
for a proton--proton collider of $8\tev$ center-of-mass energy.
However, it is also possible that new physics may be hiding in an
exotic place and we have not been ``digging'' in the right spot yet.
Therefore it is important to optimize searches based on some
particular scenarios as well as to develop model-independent search
strategies. Otherwise the new physics may be missed easily.

Among many others, searches for a new resonance (most likely originating
from a new force) are particularly important.
It may show up in an early stage of LHC Run\:2 (with low luminosity)
and it may be relatively easy to reconstruct the mass of the resonance.
Typical final states, which are searched for so far, include dijet,
dilepton, and diphoton topologies as well as signatures of $t\bar t$
pairs and so on. Current bounds on these resonances are already at
scales of a few TeV for reasonable assumptions on couplings between SM
particles and the resonance.

Especially searches for a \ttbar resonance are very well motivated and
many models contain such a resonance, such as the KK graviton in the
RS model \cite{Lillie:2007yh,Lillie:2007ve}, the level-2 KK resonances
in UED with bulk masses \cite{Kong:2010xk,Flacke:2013pla}, the coloron
\cite{Simmons:1996fz}, the axigluon \cite{Bagger:1987fz} and the
$Z^\prime$ arising from a larger gauge symmetry.
Current bounds on the \ttbar resonance are quite strong and range from
$2.0$-$2.5\tev$ for several models considered by the ATLAS and CMS
collaborations~\cite{ATLAS-CONF-2013-052,Aad:2013nca,Chatrchyan:2013lca,Chatrchyan:2012yca}.
Interestingly, in most experimental analyses, such a $t\bar{t}$
resonance is produced via an annihilation of $q\bar q$ pairs. However,
if the \ttbar resonance couples to the diquarks rather weakly or does
not couple to them at all, one should consider that this resonance
(which we label as ``$G_3$'' in this paper) may be produced in
association with another \ttbar pair \cite{Brooijmans:2010tn},
i.e.~$gg\to G_3+t\bar{t}$. This generates final states with four
(decaying) top quarks, and in this case it would be challenging to
reconstruct the resonance mass \cite{Brooijmans:2010tn}.

In this paper, we investigate the model-independent production of a
top-philic resonance.
We will assume that it dominantly couples to the top quark pair and
rarely interacts with the rest of the SM particles, i.e.~the diquark
coupling is negligible.
The resonance could be either a vector or a scalar; either a
color-octet or a color-singlet. In this study, we will focus on a
color-singlet vector particle.

We find that such a top-philic resonance can be produced in two
different ways, that is at the tree level and at the one-loop level.
The production via one loop leads to a large cross section compared to
the tree level production modes. In the absence of the diquark
coupling, the tree level production requires at least three particles
in the final state, and is therefore suppressed by phase space.
All these production modes are model-independent in a sense that they
do not rely on how the resonance couples to other particles in the model.
Even if a diquark coupling is zero, these production channels always
exist as long as one is interested in a \ttbar resonance.

A more exotic resonance, the chromophilic $Z^\prime$, is investigated
in Ref.~\cite{Alwall:2012np} and has been searched for by the CDF
collaboration~\cite{Aaltonen:2012th}. In their study, this heavy
$Z^\prime$ resonance interacts with the SM gluon only, leading to the
dominant decay mode $Z^\prime\to g^\ast g\to q\bar qg$. In our study,
we allow for an interaction of the $G_3$ to the top quark only and
assume that there is no coupling to the SM gluon. Therefore, once
produced, the $G_3$ vector particle will promptly decay into a pair of
top quarks.

\bigskip
The outline of our paper is as follows: we set up a very simple model
of a top-philic resonance and compute $G_3$ production cross sections
in Section~\ref{sec2} with emphasis on the production via one loop.
In Section~\ref{bounds}, we consider current bounds on the model
parameter space and discuss the prospects of searching for $G_3$ in
Run\:2 of the LHC. Finally, we summarize our findings in
Section~\ref{conclusion}.

\clearpage

\section{Bottom-up Approach for a Top-Philic Resonance
  ($\mathbold{G_3}$)}\label{sec2}

\subsection{The setup\label{model}}

\begin{figure}[t!]
  \centering  
  \includegraphics[width=0.67\textwidth]{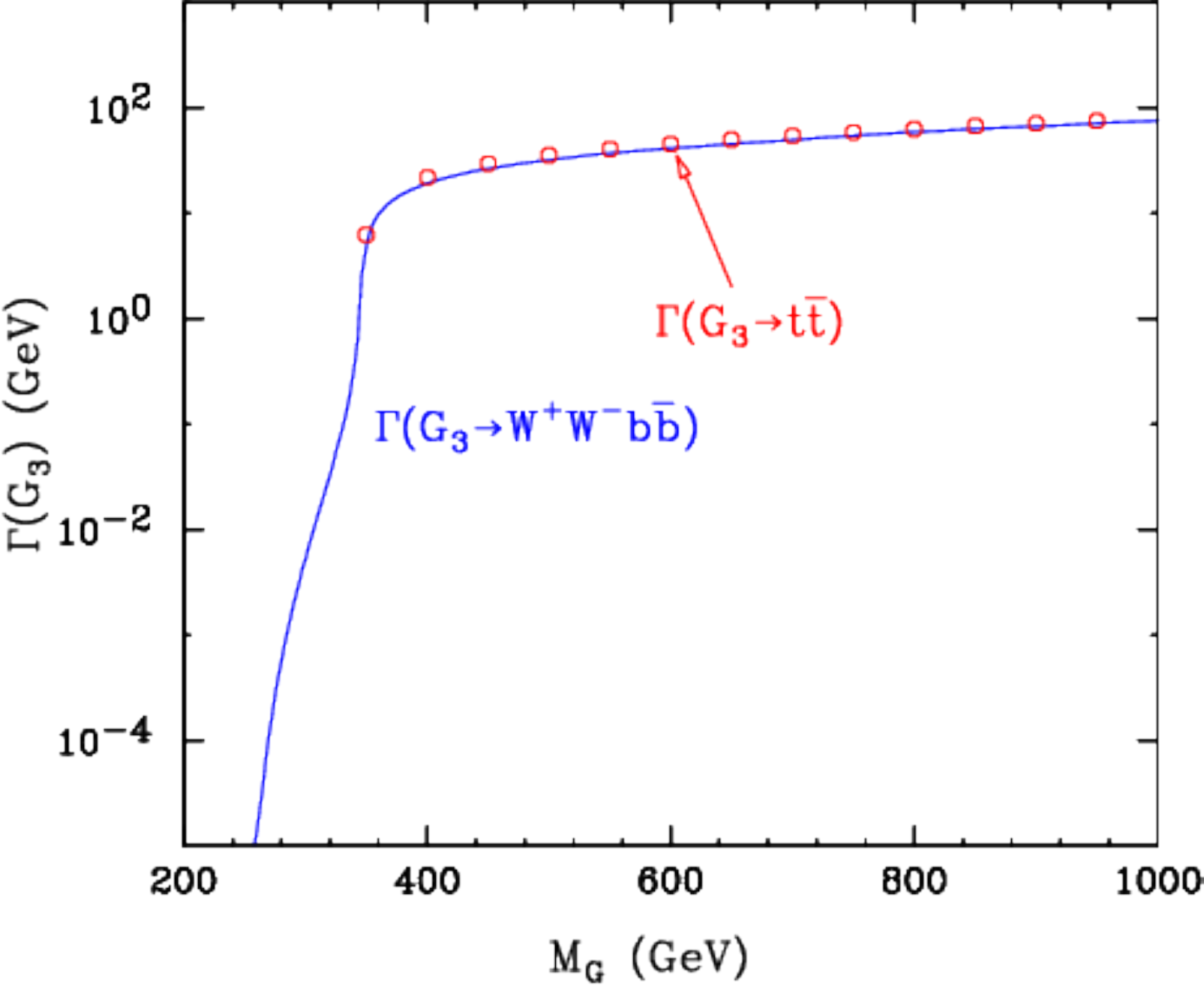}
  \caption{\label{fig:width}
    Total decay width of $G_3$. The solid curve (in blue) represents
    the results of the computation of the four-body decay
    $G_3\to W^+W^-b\bar b$, while the two-body decay yielding two
    on-shell top quarks is shown by the circles (in red).}
\end{figure}

We consider a model with a color-singlet vector particle ($G_3$) of
mass $M_G$, which couples to the top quark (\,$t$ and $\bar t$\,)
only. We assume a very weak or no interaction with all the other
quarks and the gluon. The only relevant interaction is given as
follows:
\begin{eqnarray}
{\cal L} &\;=\;&
\bar{t}\,\gamma_\mu\Big(c_L P_L + c_R P_R\Big)t\,G_3^\mu~,\nonumber\\[1mm]
         &\;=\;&
c_t\,\bar{t}\,\gamma_\mu\Big(\cos\theta\,P_L + \sin\theta\,P_R\Big)t\,G_3^\mu~,
\end{eqnarray}
where $P_{L/R}=(1\mp\gamma_5)/2$, $c_t=\sqrt{(c_L)^2+(c_R)^2}$ and
$\tan\theta=\frac{c_R}{c_L}$.
The decay width at leading order is given by 
\begin{eqnarray}
\mrm{\Gamma}\Big(G_3\to t\bar t\Big) &\;=\;&
\frac{c_t^2\,M_G}{8\pi}\;\sqrt{1-\frac{4\,m_t^2}{M_G^2}}\;
\left[1+\frac{m_t^2}{M_G^2}\,\big(3\sin2\theta-1\big)\right]~,\\[1mm]
                                     &\;\approx\;&
\frac{c_t^2\,M_G}{8\pi}\qquad\quad\mbox{for\ } m_t\ll M_G~.
\end{eqnarray}
Since $\frac{\mrm{\Gamma}}{M_G}\approx\frac{c_t^2}{8\pi}\ll 1$, the
width of the resonance is very narrow for $c_t\sim1$, and therefore
determined by the detector resolution.
Throughout this paper, we will only consider the two-body decay mode
$G_3\to t\bar t$ where $M_G > 2\,m_t$. Note that in principle the
$G_3$ can decay into a $t\,W\bar b$ system or into a $W^+b\,W^-\bar b$
system if $m_t+M_W+ m_b<M_G<2\,m_t$ or $2\,M_W+2\,m_b<M_G<m_t+M_W+m_b$,
respectively. This is shown in Figure~\ref{fig:width}. The solid curve
(in blue) has been computed considering the four-body final state
$G_3\to W^+W^-b\bar b$, while the $t\bar t$ final state with on-shell
top quarks is shown as (red) circles. Beyond the $t\bar t$ threshold,
both calculations agree.

In our setup, there are only three parameters, the mass of
the \ttbar resonance ($M_G$) and the interaction strengths,
i.e.~($c_L$, $c_R$) or ($c_t$, $\theta$). We will take a bottom-up
approach using the above Lagrangian without considering any underlying
theory. Therefore we will not assume any theoretical constraints in
our parameters and will only consider experimental bounds.
However, each underlying model may provide different constraints and
certain parameter regions may be prohibited.
For instance, a KK gluon in the RS model tends to be very heavy and
its couplings are determined by the wave function overlap in extra
dimensions.

For our model-independent approach, we find that a top-philic
resonance as advocated here can be produced in two different ways:
\begin{enumerate}
\item at the tree level where we have three main production channels
  for a $G_3$ resonance:
  \begin{enumerate}
  \item the $G_3\,t\bar t$ channel yielding a four top-quark final state,\\
    $pp\;\to\;G_3+t\bar t\;\to\;t\bar t+t\bar t$~,
  \item the $G_3\,tj$ channel yielding a three top-quark final state,\\
    $pp\;\to\;G_3+t/\bar t+j\;\to\;t\bar t+t/\bar t+j$~,
  \item and the $G_3\,tW$ channel yielding, again, a three top-quark
    final state,\\
    $pp\;\to\;G_3+t/\bar t+W^\pm\;\to\;t\bar t+t/\bar t+W^\pm$~;
  \end{enumerate}
\item at the one-loop level where we identify two production modes
  leading to di-top-quark final states, namely:
  \begin{enumerate}
  \item the loop induced $G_3\,j$ production,\\
    \quad$pp\;\to\;G_3+j\;\to\;t\bar t+j$~,
  \item and the loop induced $G_3\to t\bar t$ production from an
    off-shell $G_3$,\\
    $pp\;\to\;G_3\;\to\;t\bar t$ (off-shell)~.
  \end{enumerate}
\end{enumerate}
We will discuss all of these production channels in more detail below.

\subsection{Tree level production}\label{tree}

\begin{figure}[t!]
  \centering
  \includegraphics[width=0.32\textwidth]{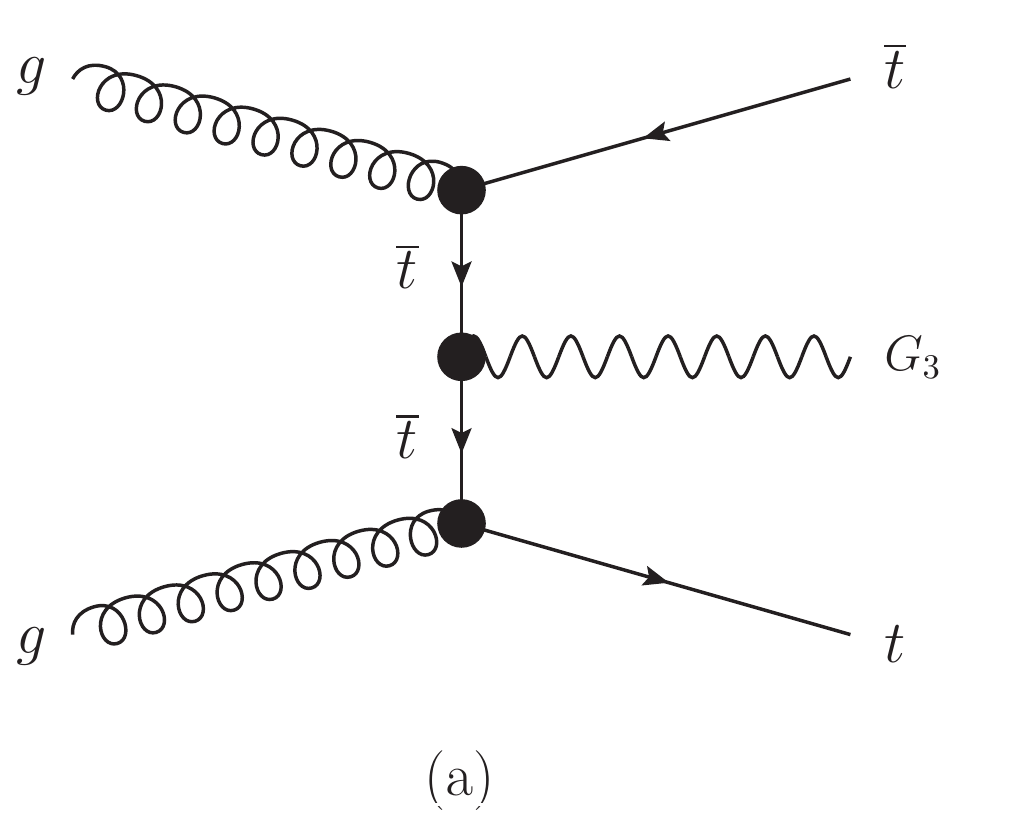}
  \includegraphics[width=0.33\textwidth,height=0.175\textheight ]{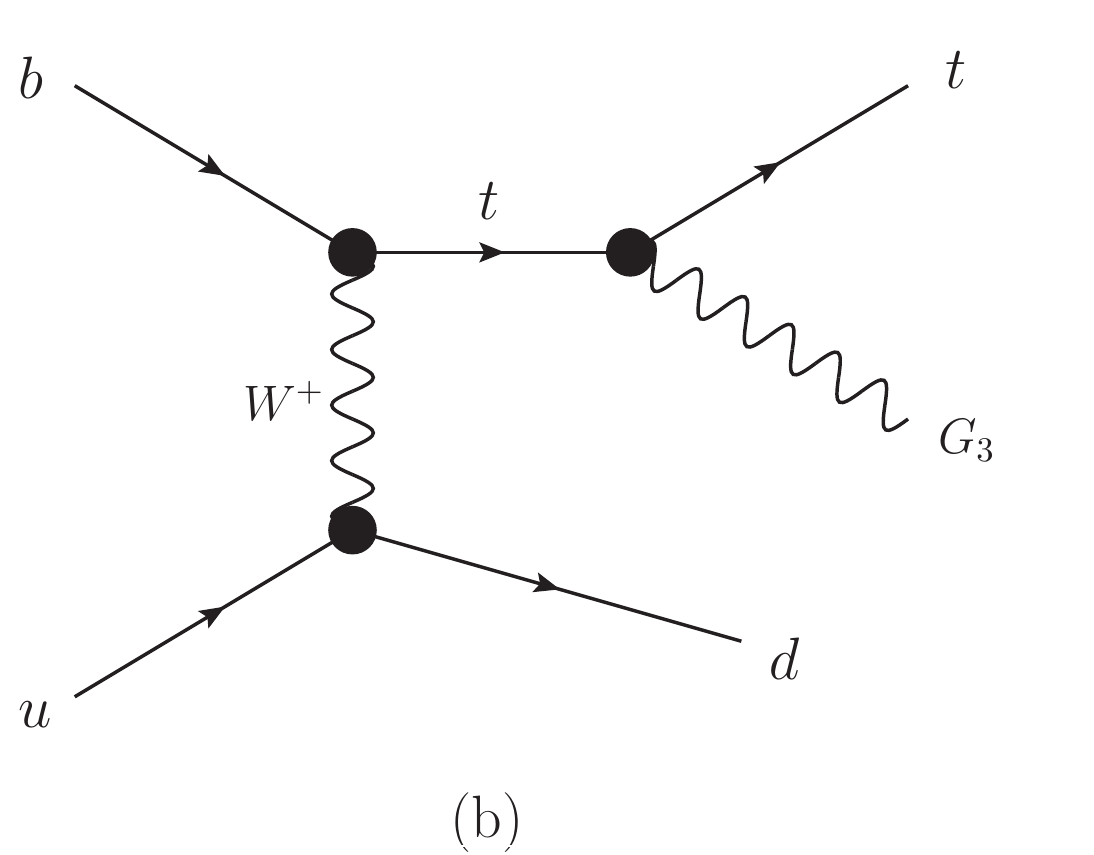}
  \includegraphics[width=0.32\textwidth]{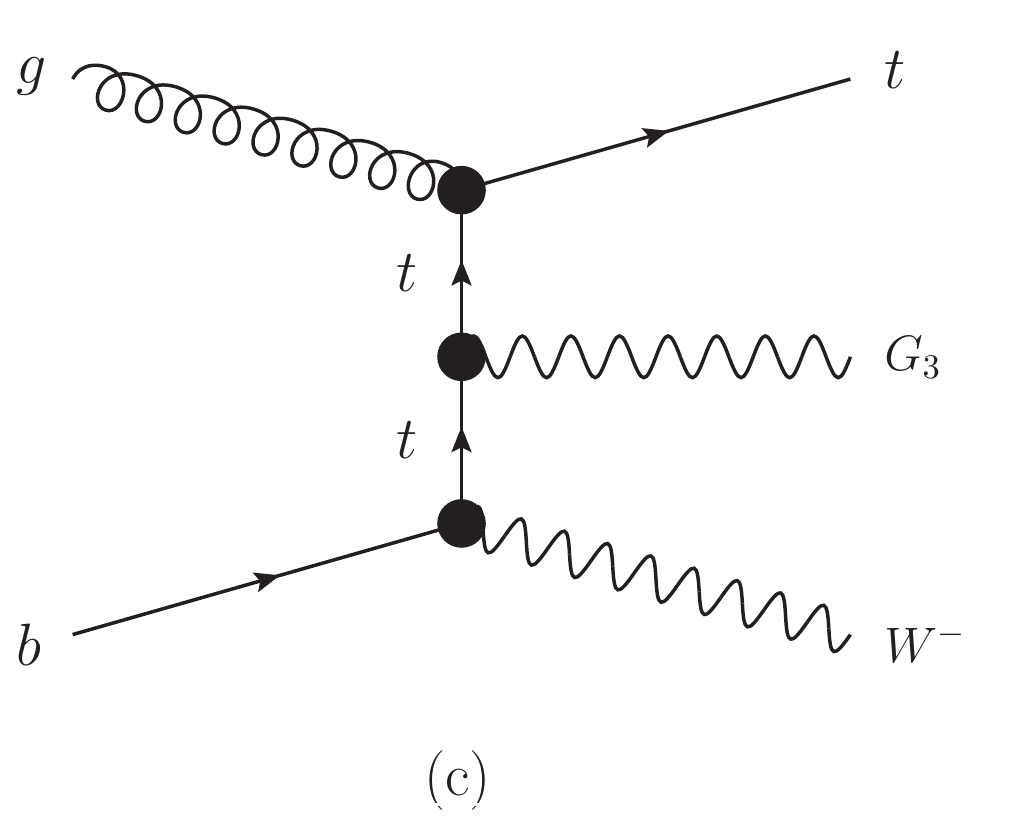}
  \caption{\label{fig:diagG3tt}
    Sample diagrams for the $G_3$ tree level production in association
    with (a) $t\bar{t}$, (b) $tj$ and (c) $tW$. The $G_3$ generation
    modes are derived from top quark final states produced via strong
    (a), electroweak (b) and mixed QCD and electroweak (c)
    interactions.}
\end{figure}

In this section, we consider the tree level production of the $G_3$
resonance, which always involves at least one extra
(i.e.~non-decay) top quark in the final state. The basic
structure of this production stems from top quark production in the
SM, see Figure~\ref{fig:diagG3tt}, where the top quark may be produced
singly ($tj/\bar tj$ and $tW^-/\bar tW^+$) or in a pair (\,$t\bar t$\,).%
\footnote{Our definition of a jet ($j$) contains the gluon and all
  five quarks ($q$) as well as their anti-particles. Note that the
  gluon ($g$) does not appear as a final state parton in the tree
  level production shown here.}
Subsequently, these top quarks in the final state radiate a $G_3$,
which will decay further into a $t\bar t$ pair. In the case of
$t\bar t$ and $tW$ production, there are additional contributions from
$\hat t$-channel radiation as shown by the left and right example
diagrams of Figure~\ref{fig:diagG3tt}, respectively.

\begin{figure}[b!]
  \centering
  \includegraphics[width=0.67\textwidth]{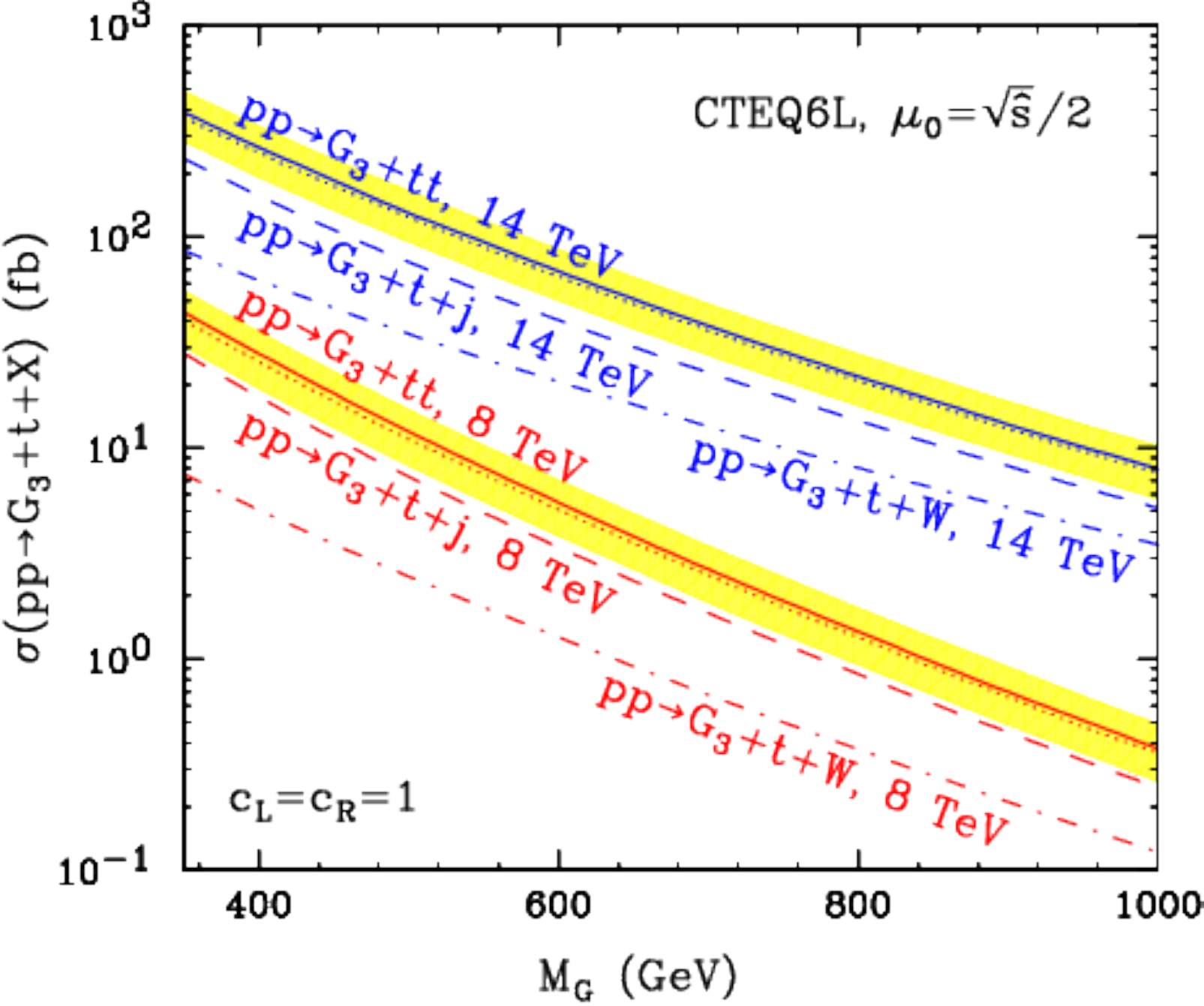}    
  \caption{\label{fig:G3}
    Tree level production cross sections of the $G_3$ vector particle
    in association with $t\bar t$ (solid), $tj$ (dashed) and $tW$
    (dot-dashed) at the LHC, using natural values for the coupling
    strengths ($c_L=c_R=1$). The cross sections are shown as a
    function of the mass of the resonance, $M_G$; the yellow bands
    correspond to scale variations by factors of two.}
\end{figure}

Figure~\ref{fig:G3} summarizes the tree level production cross
sections of the $G_3$ as a function of its mass ($M_G$) for a choice
based on natural coupling values, $c_L=1=c_R$. We use
\textsc{CalcHEP}~\cite{Belyaev:2012qa} for the cross section
evaluation and verified the results using
\textsc{MadGraph5}~\cite{Alwall:2011uj}.
We set the renormalization and factorization scale to
$\mu_0=\sqrt{\hat s}/2$, and employ the CTEQ6L parton density
functions for all computations. Especially for the $G_3\,t\bar t$
production, this scale choice gives consistent results compared to the
one reading $\mu_0=M_G/2+m_t$, which is similar to $M_h/2+m_t$ used in
$t\bar{t}h$ production. The cross section results for the processes
$pp\to G_3+t\bar{t}$, $pp\to G_3+tj$ and $pp\to G_3+tW$ are shown as
solid, dashed, and dot-dashed curves, respectively, with the $8\tev$
results depicted in blue and the $14\tev$ results depicted in red. No
cuts have been imposed in Figure~\ref{fig:G3} and all cross sections
are compared at the parton level.

\begin{figure}[t!]
  \centering
  \begin{subfigure}[b]{0.47\textwidth}\centering
    \includegraphics[width=1\textwidth]{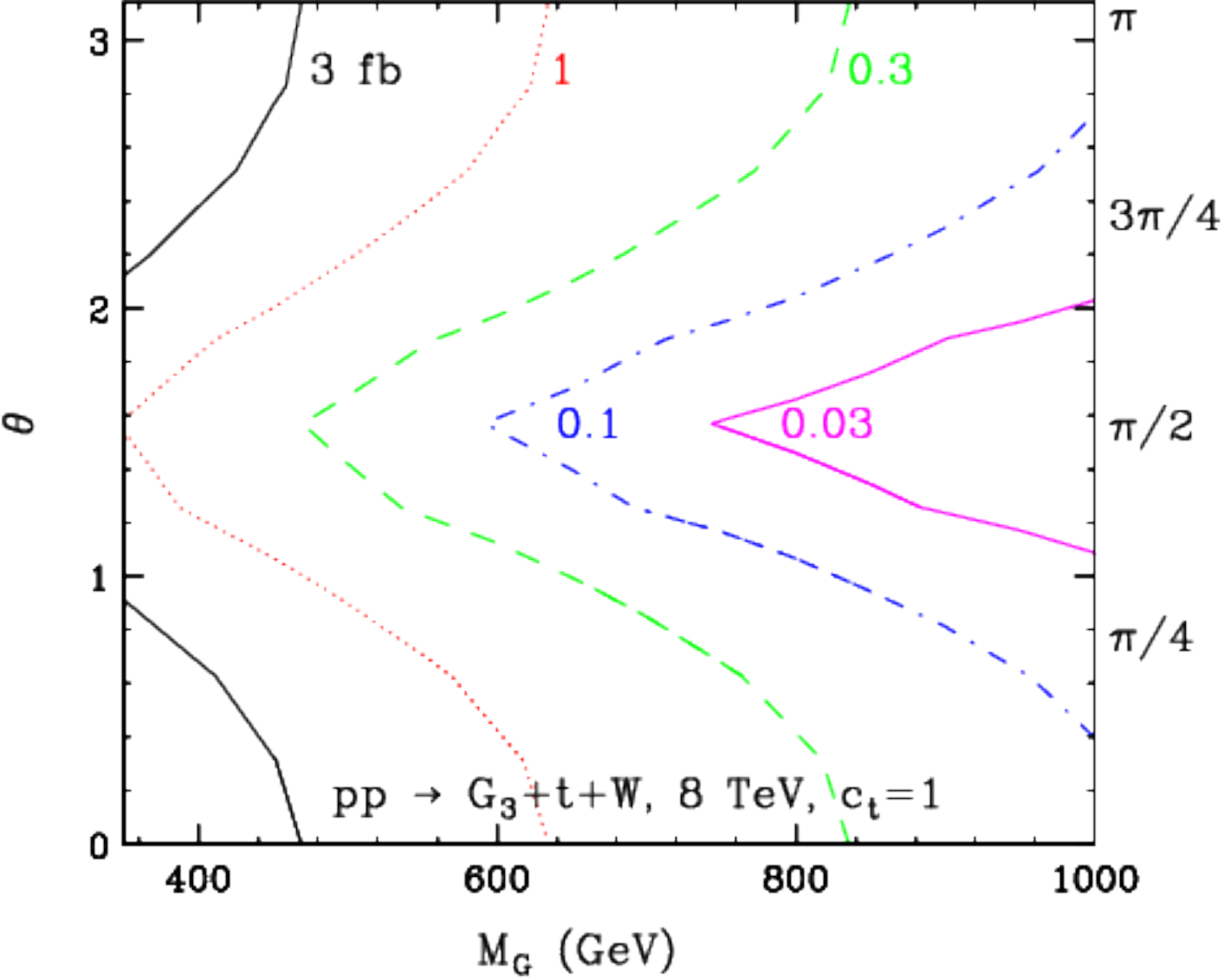}
    \caption{\label{sfig:g3tW8}}
  \end{subfigure}
  \hskip5mm
  \begin{subfigure}[b]{0.47\textwidth}\centering
    \includegraphics[width=1\textwidth]{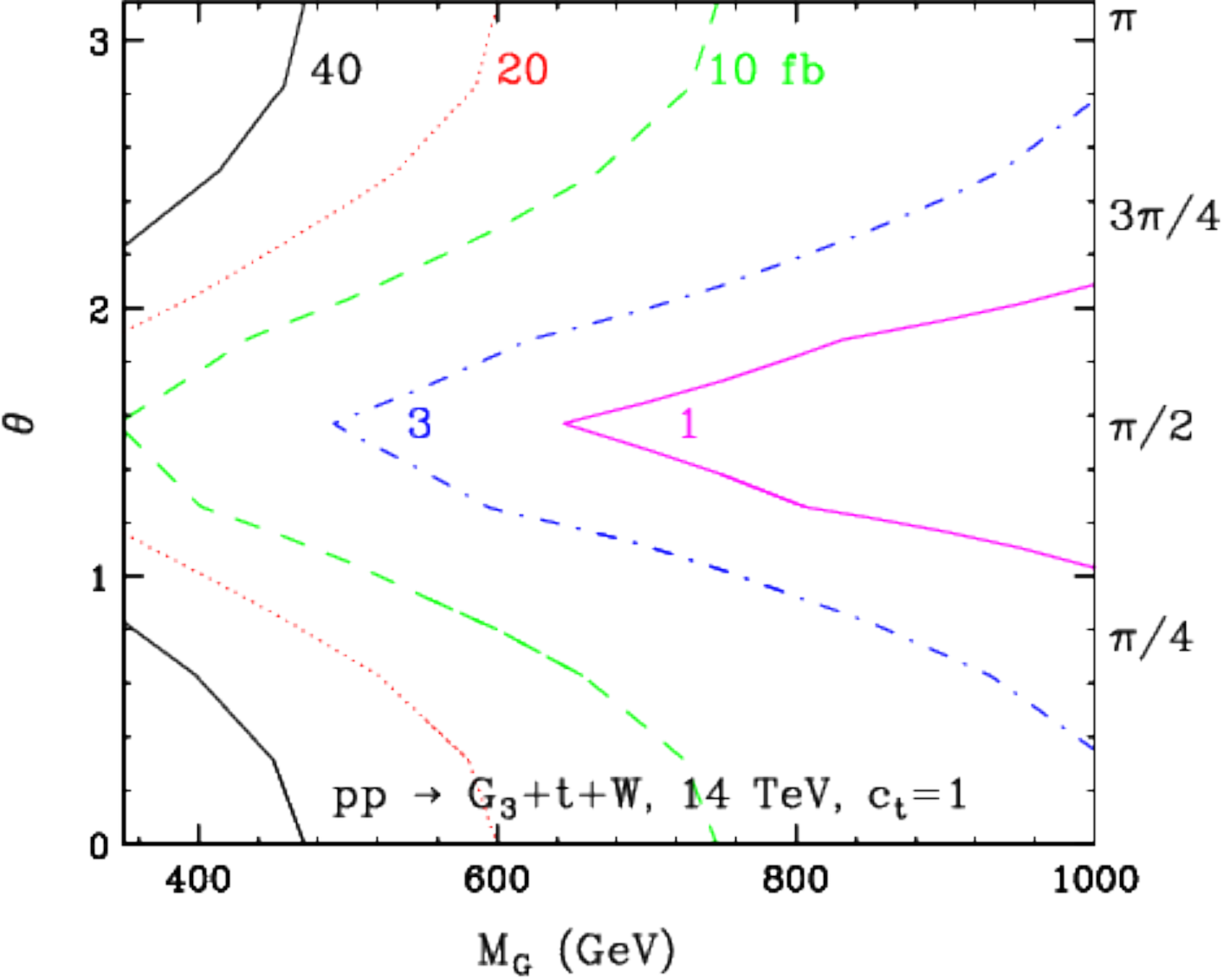}
    \caption{\label{sfig:g3tW14}}
  \end{subfigure}
  \caption{\label{fig:g3tW}
    Tree level production cross section of $pp\to G_3+tW$ at $8\tev$
    (\ref{sfig:g3tW8}) and at $14\tev$ (\ref{sfig:g3tW14}) in the
    $M_G$-$\theta$ plane of model parameters using $c_t=1$.}
\end{figure}

\begin{figure}[t!]
  \centering
  \begin{subfigure}[b]{0.47\textwidth}\centering
    \includegraphics[width=1\textwidth]{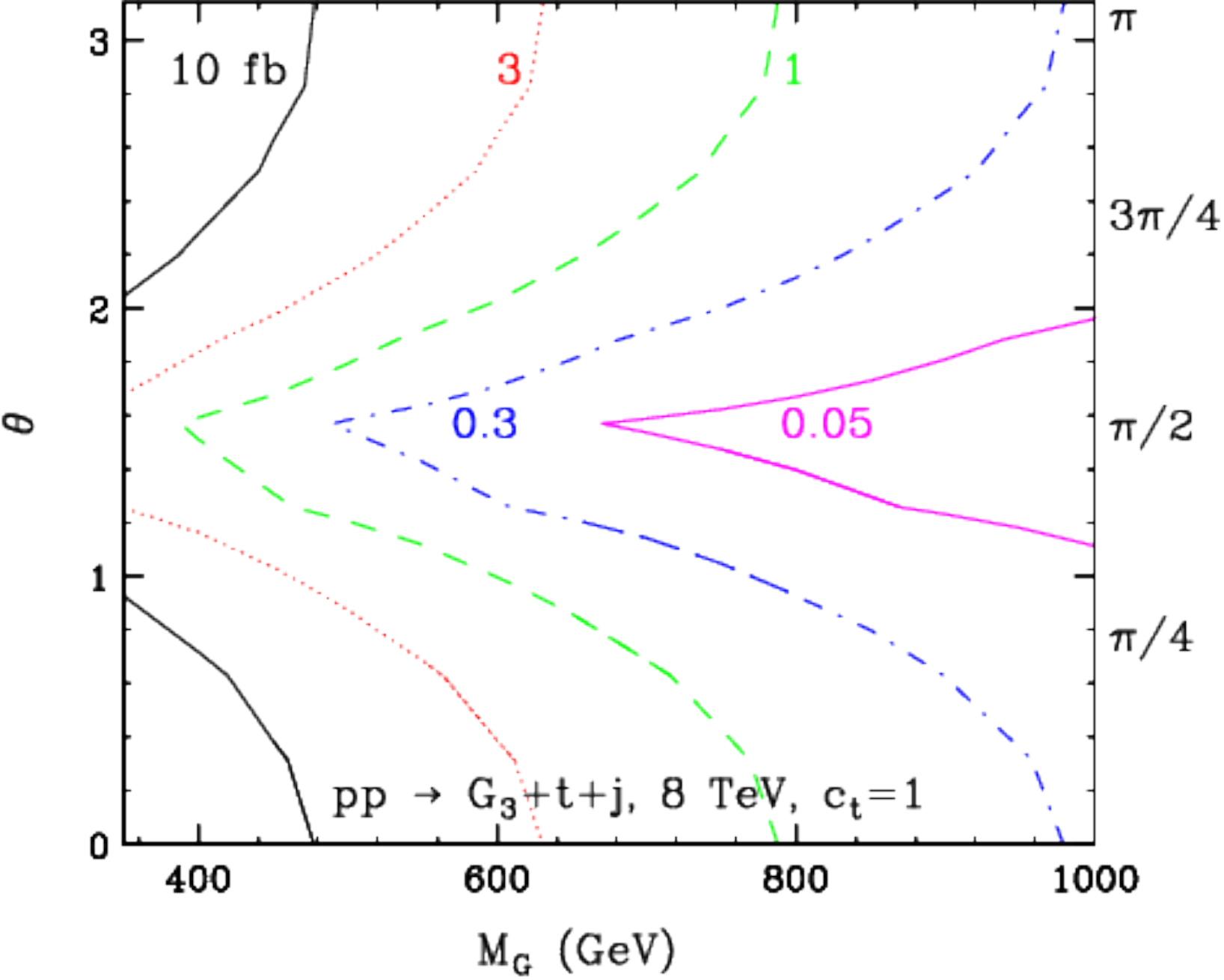}
    \caption{\label{sfig:g3tj8}}
  \end{subfigure}
  \hskip5mm
  \begin{subfigure}[b]{0.47\textwidth}\centering
    \includegraphics[width=1\textwidth]{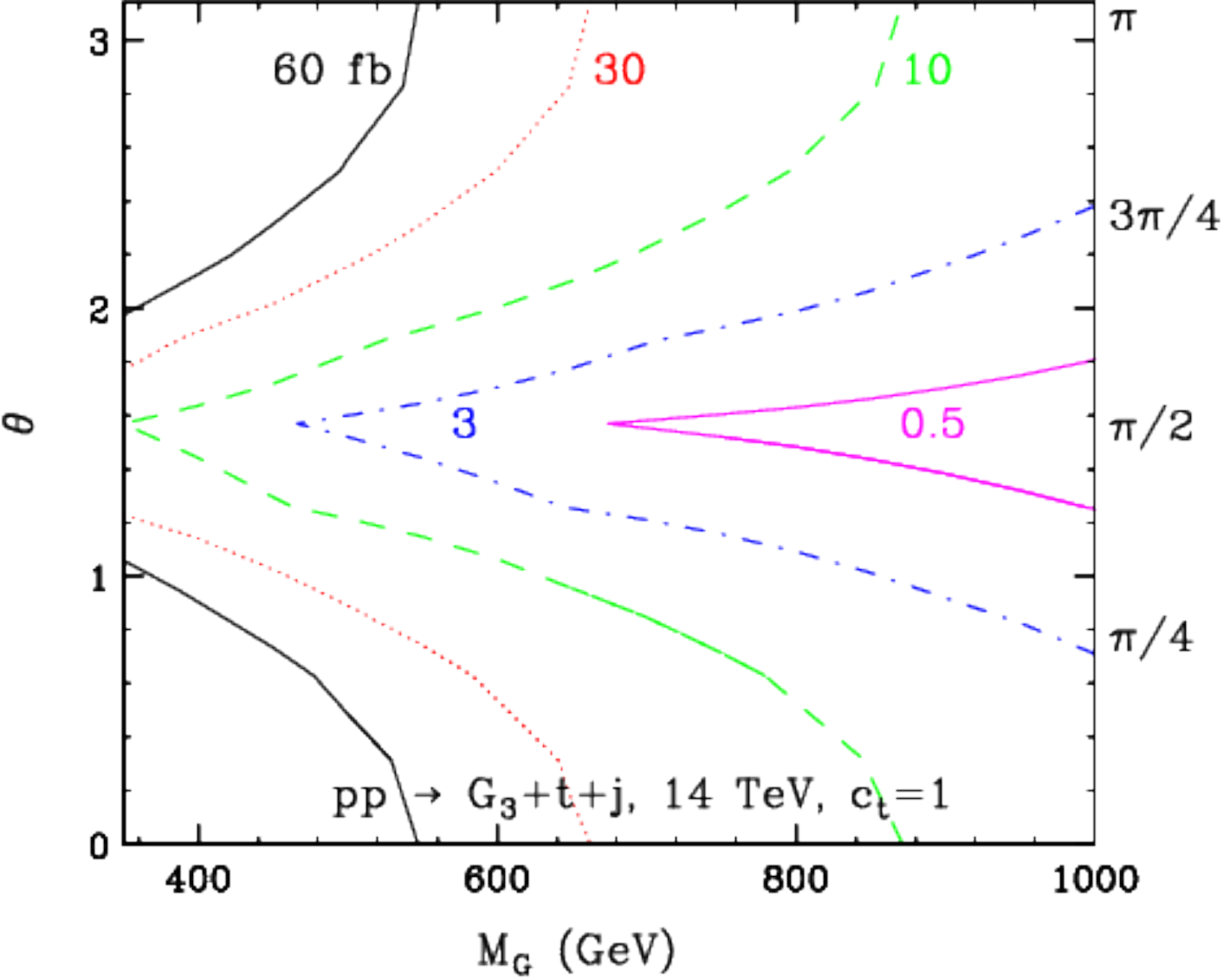}
    \caption{\label{sfig:g3tj14}}
  \end{subfigure}
  \caption{\label{fig:g3tj}
    Tree level production cross section of $pp\to G_3+tj$ at $8\tev$
    (\ref{sfig:g3tj8}) and at $14\tev$ (\ref{sfig:g3tj14}) in the
    $M_G$-$\theta$ plane of model parameters using $c_t=1$.}
\end{figure}

For $pp\to G_3+t\bar t$ (solid lines in Figure~\ref{fig:G3}), there
are 112 diagrams (including electroweak processes such as
$q\bar q\to Z\to t\bar t+G_3$). Among these there
are 8 diagrams associated with the $gg\to G_3+t\bar t$ subchannel,
which turns out to give the dominant contribution as shown by the
dotted curves in Figure~\ref{fig:G3}.
The gluon initial state contribution accounts for about $94\%$
($90$-$95\%$) of the total production cross section at $14\tev$
($8\tev$). The yellow band represents the uncertainty that arises by
varying the scales between $\mu_0/2$ and $2\,\mu_0$.
Four top-quark final states arise in many models beyond the SM.
Examples include the top compositeness model~\cite{Kumar:2009vs,Lillie:2007hd},
the topcolor-assisted technicolor model~\cite{Han:2012qu}, and models
that produce a pair of $t\bar t$ resonances such as the axigluon, the
KK gluon~\cite{AguilarSaavedra:2011ck}, the color-octet
scalar~\cite{Plehn:2008ae,Dobrescu:2007yp} and the color-sextet
scalar~\cite{Chen:2008hh}.
The production of four top-quark final states, which we propose here,
is similar to that of $t\bar tH$, where $H$ is a heavy neutral Higgs
boson, which decays to a top quark pair.

The electroweak production of a $G_3+tj$ (dashed lines in
Figure~\ref{fig:G3}) is mediated by a $W$ boson as exemplified in the
middle diagram in Figure~\ref{fig:diagG3tt}. The major part of the
cross section comes from the initial states with a $b$-quark, which
can be understood in terms of the PDF and CKM structure. Similarly,
the $g$ plus $b$-quark initial states provide the dominant contribution
to the $G_3\,tW$ production (dot-dashed lines in Figure~\ref{fig:G3}).

The $G_3\,tW$ production is the smallest owing to the suppression from
the three-body phase space and the heaviness of the final state
particles. $G_3\,tj$ is slightly larger than $G_3\,tW$ due to the jet
multiplicity and the lightness of jets compared to the $W$ boson. The
$G_3\,t\bar t$ channel gives the largest cross section due to the fact
that the top quark pair production is mainly governed by the strong
interaction, overpowering the effect from the phase-space suppression.

\begin{figure}[t!]
  \centering
  \begin{subfigure}[b]{0.47\textwidth}\centering
    \includegraphics[width=1\textwidth]{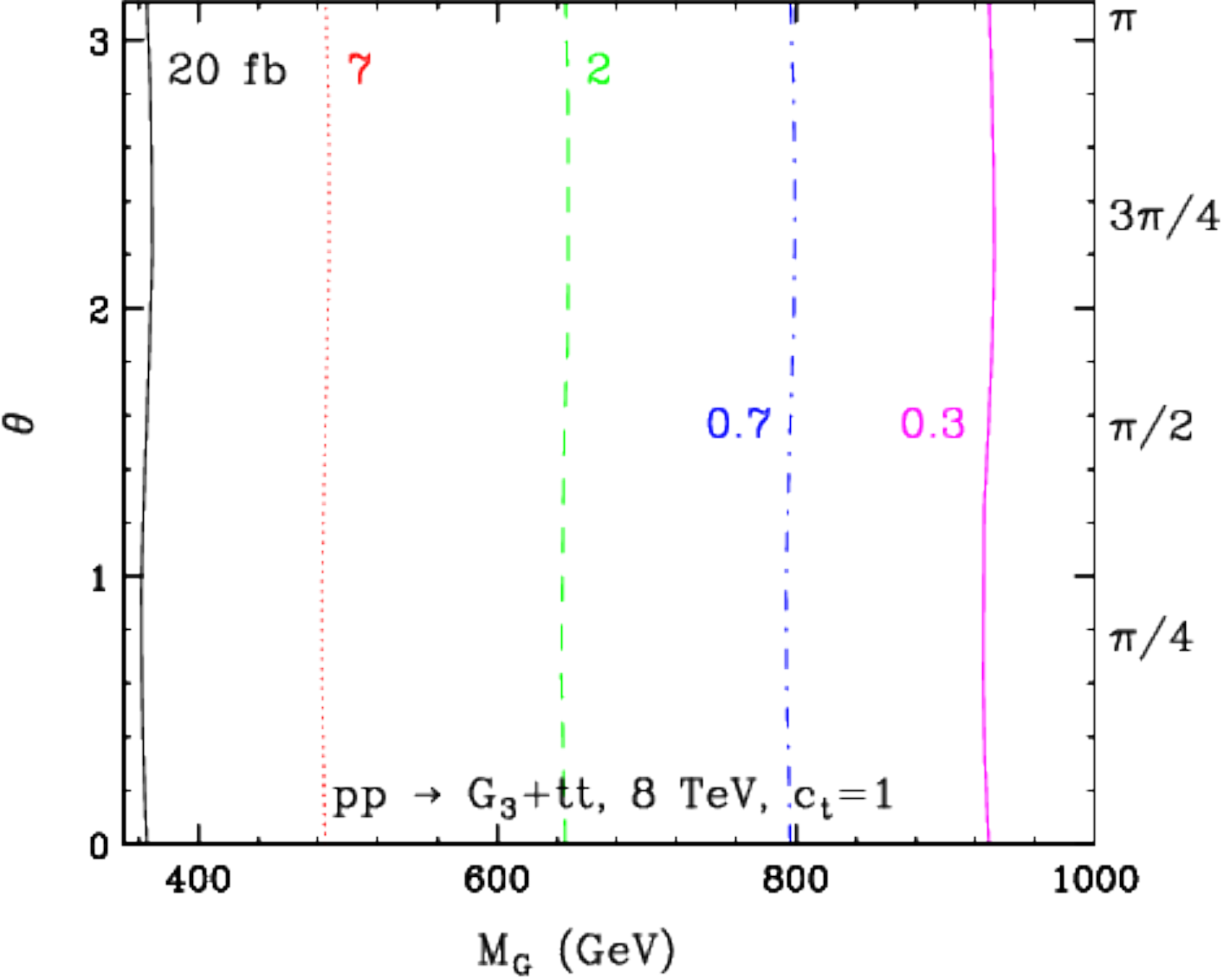}
    \caption{\label{sfig:g3tt8}}
  \end{subfigure}
  \hskip5mm
  \begin{subfigure}[b]{0.47\textwidth}\centering
    \includegraphics[width=1\textwidth]{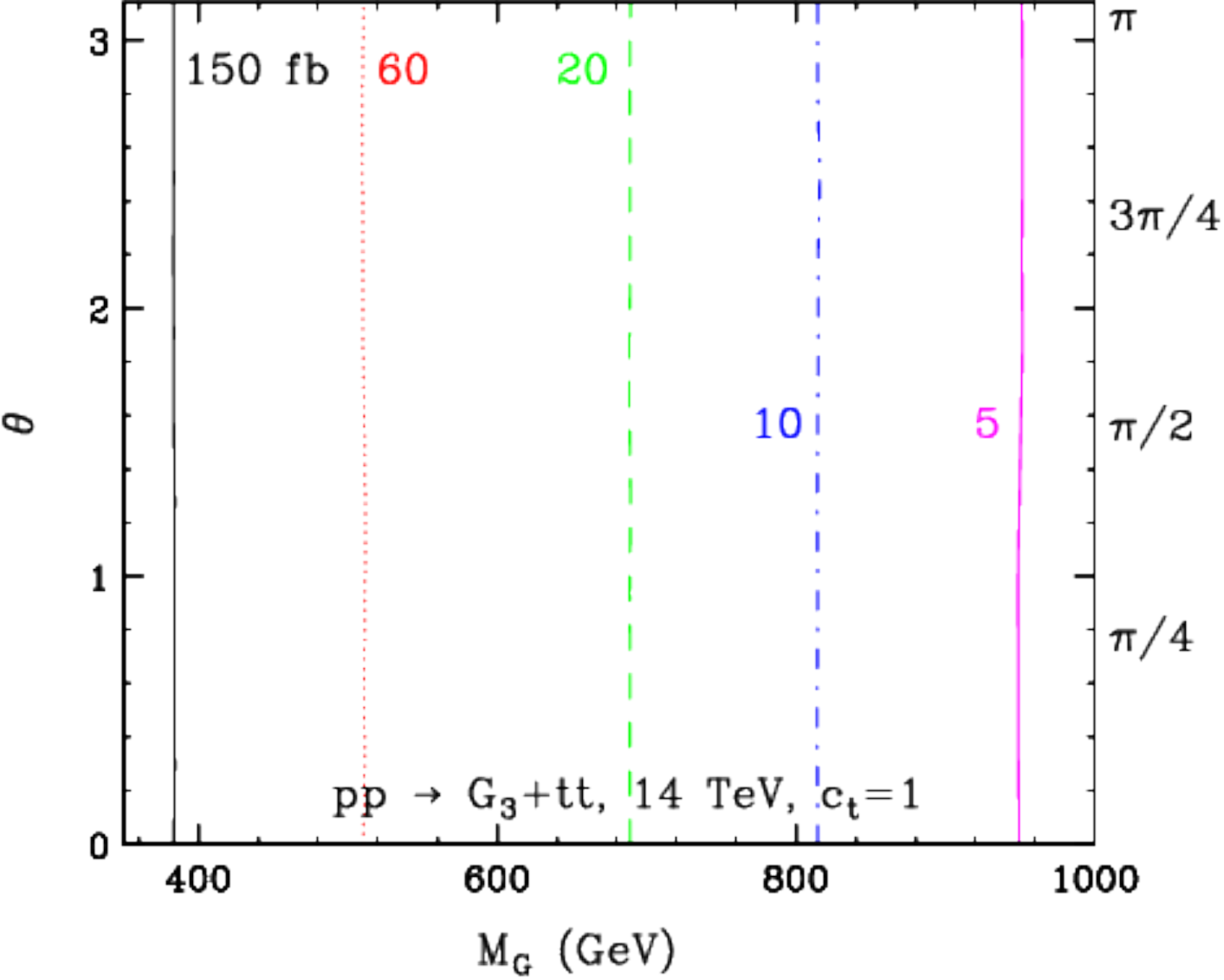}
    \caption{\label{sfig:g3tt14}}
  \end{subfigure}
  \caption{\label{fig:g3tt}
    Tree level production cross section of $pp\to G_3+t\bar t$ at
    $8\tev$ (\ref{sfig:g3tt8}) and at $14\tev$ (\ref{sfig:g3tt14}) in
    the $M_G$-$\theta$ plane of model parameters using $c_t=1$.}
\end{figure}

We find that the $G_3$ production associated with a single top quark
exhibits a strong dependence on the chirality as shown in
Figures~\ref{fig:g3tW} and~\ref{fig:g3tj}, while the $G_3$ production
accompanied by a top quark pair does not change as a function of
chirality (see Figure~\ref{fig:g3tt}). The difference between these
two findings stems from the fact that $G_3\,tW$ and $G_3\,tj$
production is mediated by the $t$-$W$-$b$ interaction. For these
processes, $\theta=0$ ($\theta=\pi/2$) corresponds to the pure
left-handed (right-handed) interaction and leads to the largest
(smallest) cross section for a given mass of the $G_3$. On the
contrary, the $G_3\,t\bar t$ process does not possess a $t$-$W$-$b$
coupling and we checked analytically that in this case the dependence
on the chirality angle $\theta$ drops out. Note that the cross
sections (based on $c_t=1$) in Figures~\ref{fig:g3tW}-\ref{fig:g3tt}
are consistent with the results shown in Figure~\ref{fig:G3}, taking
the change from $c_t=1\to c_t=\sqrt 2$ into account.

\begin{figure}[t!]
  \centering
  \includegraphics[width=0.32\textwidth]{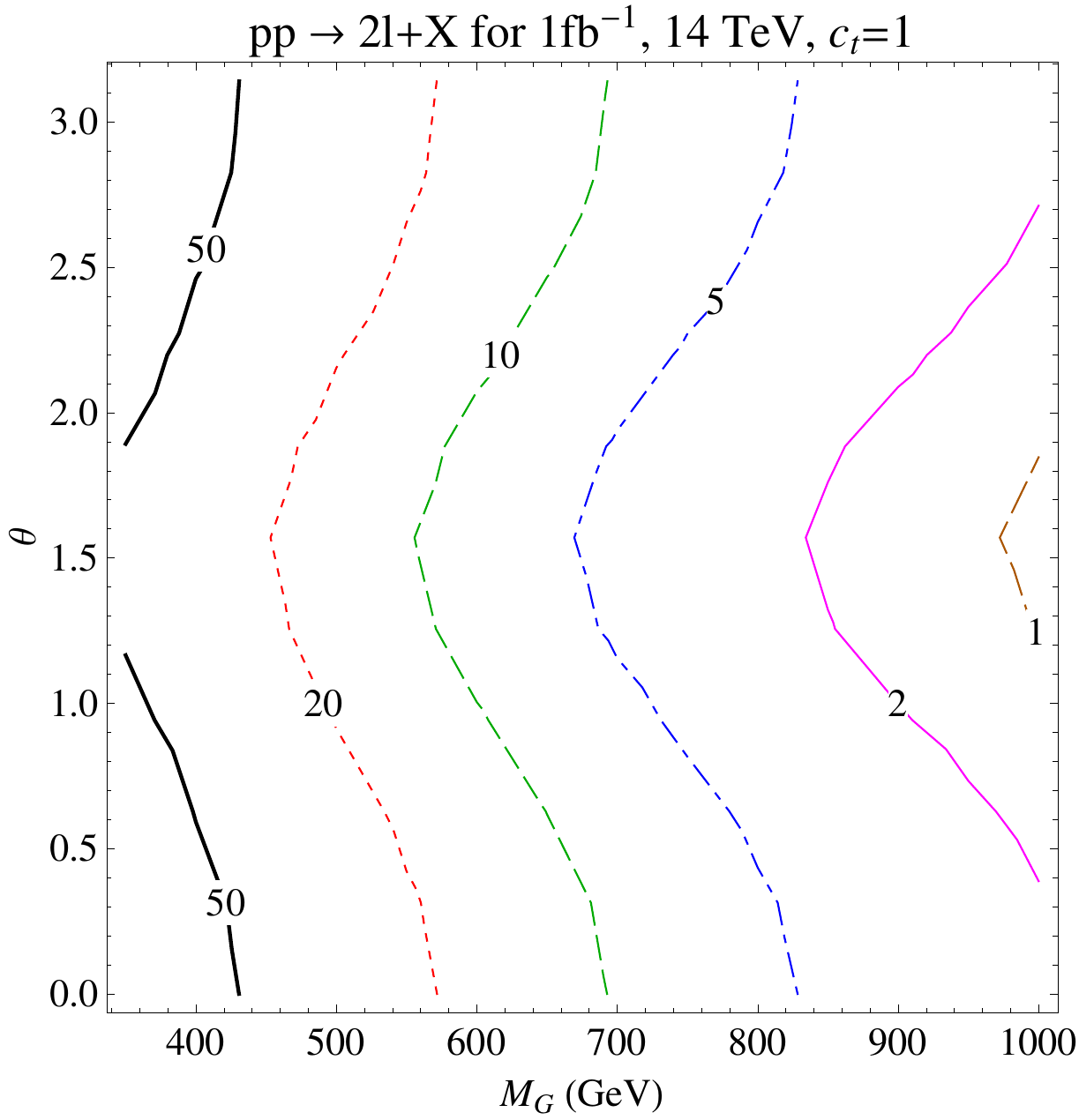}
  \hskip1mm
  \includegraphics[width=0.32\textwidth]{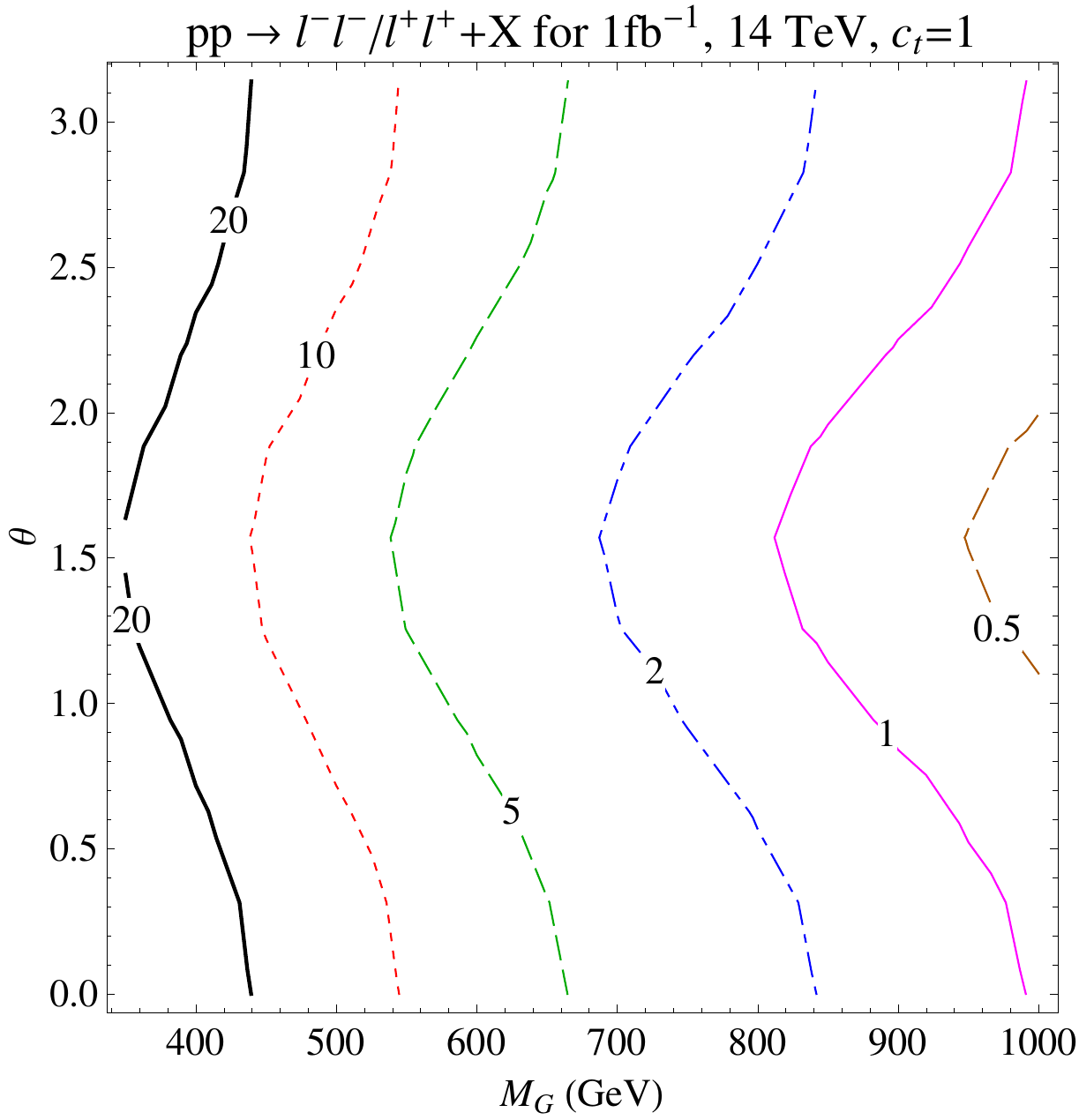}
  \hskip1mm
  \includegraphics[width=0.32\textwidth]{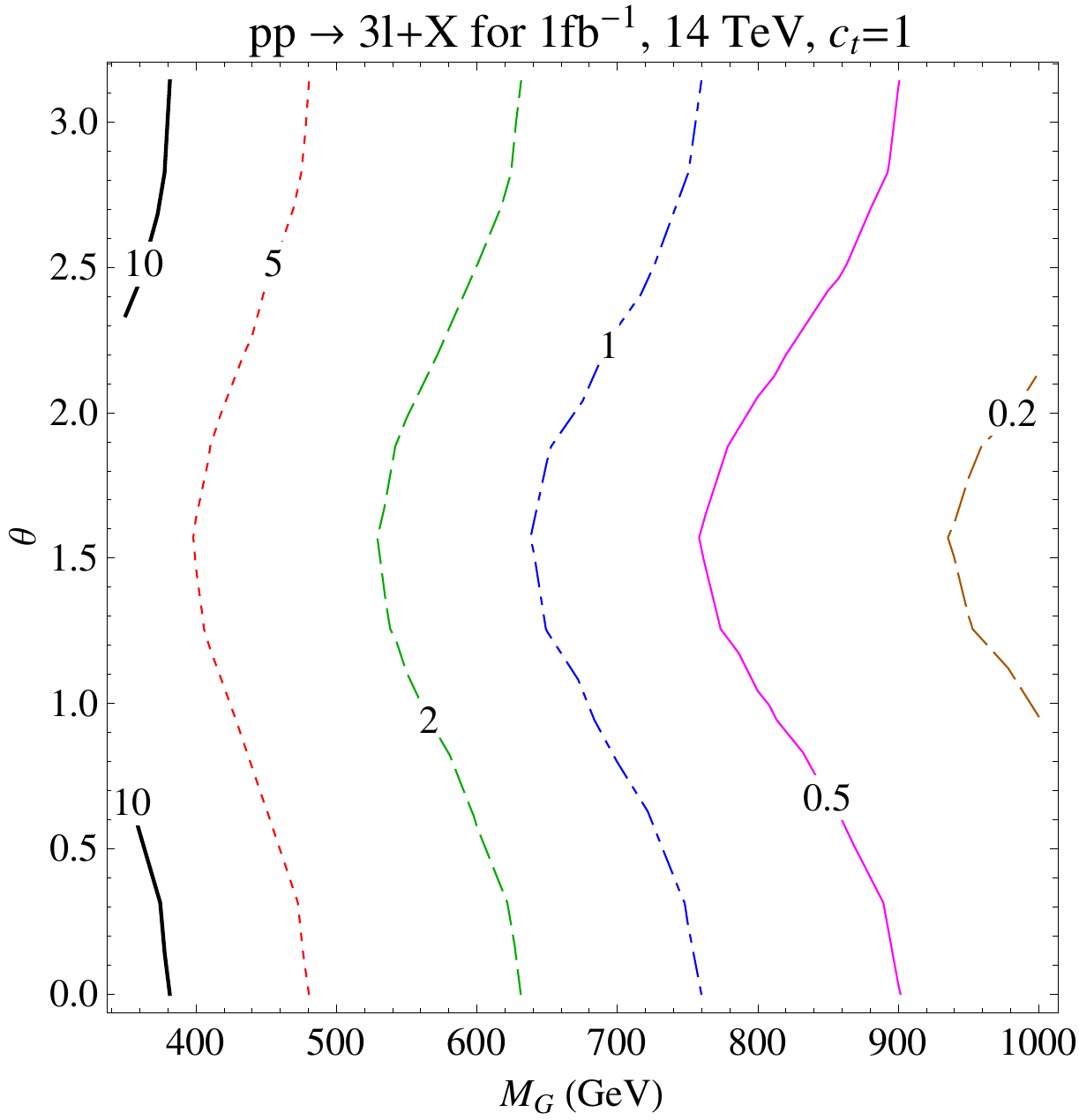}
  \caption{\label{fig:leptons}
    The expected number of events from decays of tree level generated
    $G_3$ resonances that contribute to final states for the inclusive
    production of dileptons (left), same-sign dileptons (middle), and
    trileptons (right).}
\end{figure}

In the literature, only $G_3\,t\bar t$ production is studied owing to
its large cross section but we stress that the cross section for
$G_3\,tj$ is smaller only by a factor of two and it provides a good
opportunity at the $14\tev$ LHC. Especially for resonance structure
searches in an inclusive channel such as in multi-lepton final states,
all production modes contribute at a non-negligible rate. The
inclusive search is important as the tree level production leads to
multi-gauge boson (or multi-top quark) final states, which make it
difficult to reconstruct a bump in a straightforward manner.
Figure~\ref{fig:leptons} shows the number of expected signal events
in the inclusive dilepton, same-sign dilepton and trilepton final
state. The cross sections have been computed for $c_t=1$ and are
normalized to $1\fbi$ at $14\tev$. The four-lepton events are roughly
$5$-$6\%$ of the events that have at least three leptons in the final
state. As shown in Figure~\ref{fig:leptons}, the same-sign dilepton
events are fairly large and make up a fraction of about $40$-$45\%$ of
the inclusive dilepton events. The statistics for dilepton events is
higher than for trilepton events by a factor of 7 but a larger
background is expected for the dilepton selection. Note that all
curves in Figure~\ref{fig:leptons} are just rough estimates of such
event counts, since no cuts have been imposed in this study.

We finally remark that the above discussion of the tree level
generation of the $G_3$ resonance has been guided by the principle of
modifying resonant SM top quark topologies in single top quark
(electroweak interaction) and top quark pair (strong interaction)
production by permitting the real emission of a $G_3$ particle off one
of the top quark lines. However, as one implication of a more complete
physics picture we immediately realize that the $G_3$ can also emerge
from SM topologies with non-resonant top quarks. For example, in the
$gg\to bW^+W^-\bar b$ process, the $G_3$ can be attached to a
$\hat t$-channel top quark propagator connecting two ingoing
$b$-quarks each of which originating from an initial state gluon
splitting and emitting a $W$ boson subsequently. At the inclusive
level -- which we discuss in this paper -- non-resonant contributions
like the one in this example are usually small and can be neglected.
However, one should bear in mind that more sophisticated kinematic
selections project out phase-space regions where these contributions
may become relatively large. Furthermore, a $G_3$ emission can shift
the top quark propagator onto its mass-shell. In these
cases, it is important to go beyond the approximation of factorizing
the top quark and $G_3$ production from their decays. For instance,
the $G_3\,tW$ and $G_3\,t\bar t$ are the singly and doubly top quark
resonant parts of just one, more inclusive, more physical,
$G_3\,W^+bW^-\bar b$ calculation (in the four-flavour scheme). This is
very similar to realizing that the $tW$ and $t\bar t$ configurations
are contained in the calculation of the one SM process
$pp\to WWb\bar b$ in the four-flavour scheme.

\subsection{Loop induced production}\label{loop}

In this section, we focus on the new class of production modes, which,
as mentioned before, only arise at the one-loop level.

The virtual amplitudes at one loop have been generated with
\GOSAM \cite{Cullen:2014yla,Cullen:2011ac}, a publicly available
package for the automated generation of one-loop amplitudes. It is
based on a Feynman diagrammatic approach using
{\textsc{QGRAF}}~\cite{Nogueira:1991ex} and
{\textsc{FORM}}~\cite{Vermaseren:2000nd} for the diagram generation, and
{\textsc{Spinney}}~\cite{Cullen:2010jv},
{\textsc{Haggies}}~\cite{Reiter:2009ts} and
{\textsc{FORM}} to write an optimized Fortran output. For the
reduction of the tensor integrals, we used
{\textsc{Ninja}}~\cite{Mastrolia:2012bu,vanDeurzen:2013saa,Peraro:2014cba},
an automated package for the integrand reduction via Laurent
expansion. This package is a part of \GOSAM and therefore no further
work is required to use it. Alternatively, one can use other reduction
techniques such as integrand reduction using the OPP
method~\cite{Ossola:2006us,Mastrolia:2008jb,Ossola:2008xq} as
implemented in
{\textsc{Samurai}}~\cite{Mastrolia:2010nb} or methods of tensor
reduction as contained in
{\textsc{Golem95}}~\cite{Heinrich:2010ax,Binoth:2008uq,Cullen:2011kv,Guillet:2013msa}.
The remaining scalar integrals have been evaluated using
{\textsc{OneLoop}}~\cite{vanHameren:2010cp}.

We can obtain the diagrams involving a $G_3$ resonance by using a SM
$Z$ boson, require that the $Z$ boson couples only to top quarks, and
modify its parameters such as the mass, the width, and the couplings
to vector- and axial-currents accordingly.
All diagrams for all subprocesses and production modes are ultraviolet
and infrared finite, i.e.~all possible double and single
poles of the one-loop amplitude are zero. In \GOSAM this ``zero'' is
obtained numerically which means deviations
from zero can be used to assess possible numerical
instabilities. Therefore we checked the pole terms
for each phase-space point and any event was rejected if the pole
terms were in agreement with zero
in less than thirteen digits. The fraction of such events was however
in the sub per-mill range which
indicates a numerically stable evaluation of the amplitudes.

\begin{figure}[t!]
  \centering
  \begin{subfigure}[b]{0.4\textwidth}\centering
    \includegraphics[width=1\textwidth]{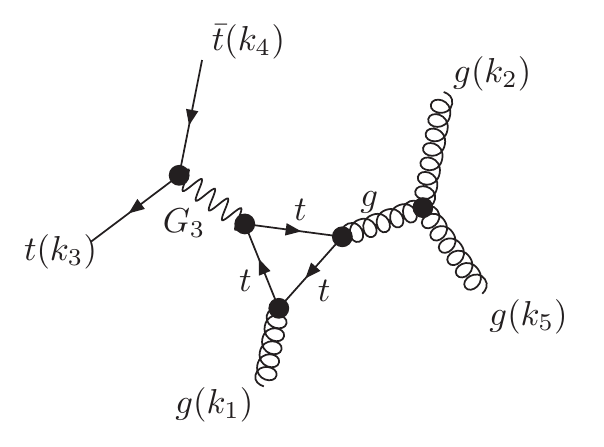}
    \caption{\label{sfig:diagsttj1}}
  \end{subfigure}
  \hskip5mm
  \begin{subfigure}[b]{0.4\textwidth}\centering
    \includegraphics[width=1\textwidth]{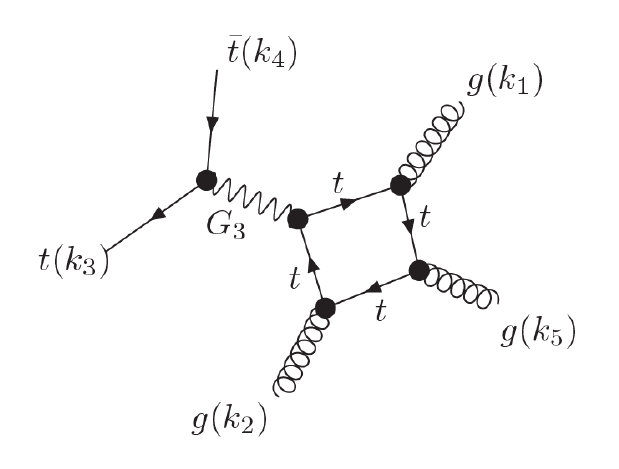}
    \caption{\label{sfig:diagsttj2}}
  \end{subfigure}
  \\[5mm]
  \begin{subfigure}[b]{0.4\textwidth}\centering
    \includegraphics[width=1\textwidth]{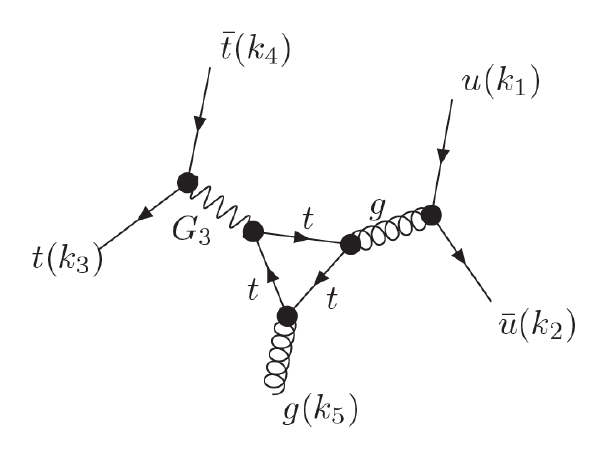}
    \caption{\label{sfig:diagsttj3}}
  \end{subfigure}
  \hskip5mm
  \begin{subfigure}[b]{0.4\textwidth}\centering
    \includegraphics[width=1\textwidth]{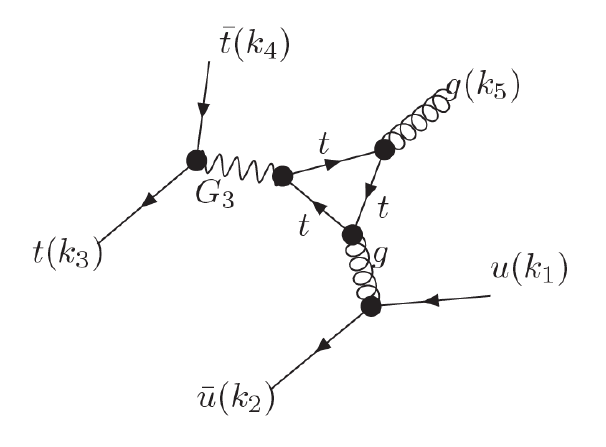}
    \caption{\label{sfig:diagsttj4}}
  \end{subfigure}
  \caption{\label{diagsttj}
    Loop induced production mode of the $G_3\,(\to t\bar t)$ in
    association with a jet: sample diagrams for the gluon--gluon
    initial state and the two single diagrams that contribute to the
    quark initiated subprocess.}
\end{figure}

For the numerical integration over the phase space, we used
\textsc{MadEvent} \cite{Maltoni:2002qb}. To improve on the timing for
the evaluation of a phase-space point, we introduced a Monte Carlo
sampling over the helicities for the gluon initiated subprocess in the
$t\bar t+\mbox{jet}$ channel (see Section~\ref{associated}).

For the theoretical predictions, we used the following setup and
parameters:
\begin{equation}
  p_{T,j}\;>\;25\gev\,,\quad |\eta_j|\;<\;2.5\,,\quad
  R\;=\;0.4\,,       \quad m_t\;=\;173.4\gev\,,\quad
  \mrm{\Gamma}_t\;=\;1.5\text{\:GeV}~.
  \label{eq:setup}
\end{equation}
Both the renormalization and factorization scale were set to
$\mu_\mbx{R,F}=\sqrt{\hat s}/2$, and we used the CTEQ6L PDF set.
Concerning the jet definition, this setup defines in particular what
we will refer to as the experimental cuts. Later on, we will also make
use of the loose cuts regime where we relax the requirements on the
jets reading
\begin{equation}
  p_{T,j}\;>\;20\gev
  \qquad\mbox{and}\qquad
  |\eta_j|\;<\;6.0~.
  \label{eq:setup.loose}
\end{equation}

\boldmath
\subsubsection{Associated production: $pp\to G_3+j\to t\bar t+j$}
\unboldmath\label{associated}

\begin{figure}[t!]
  \centering
  \includegraphics[width=0.67\textwidth]{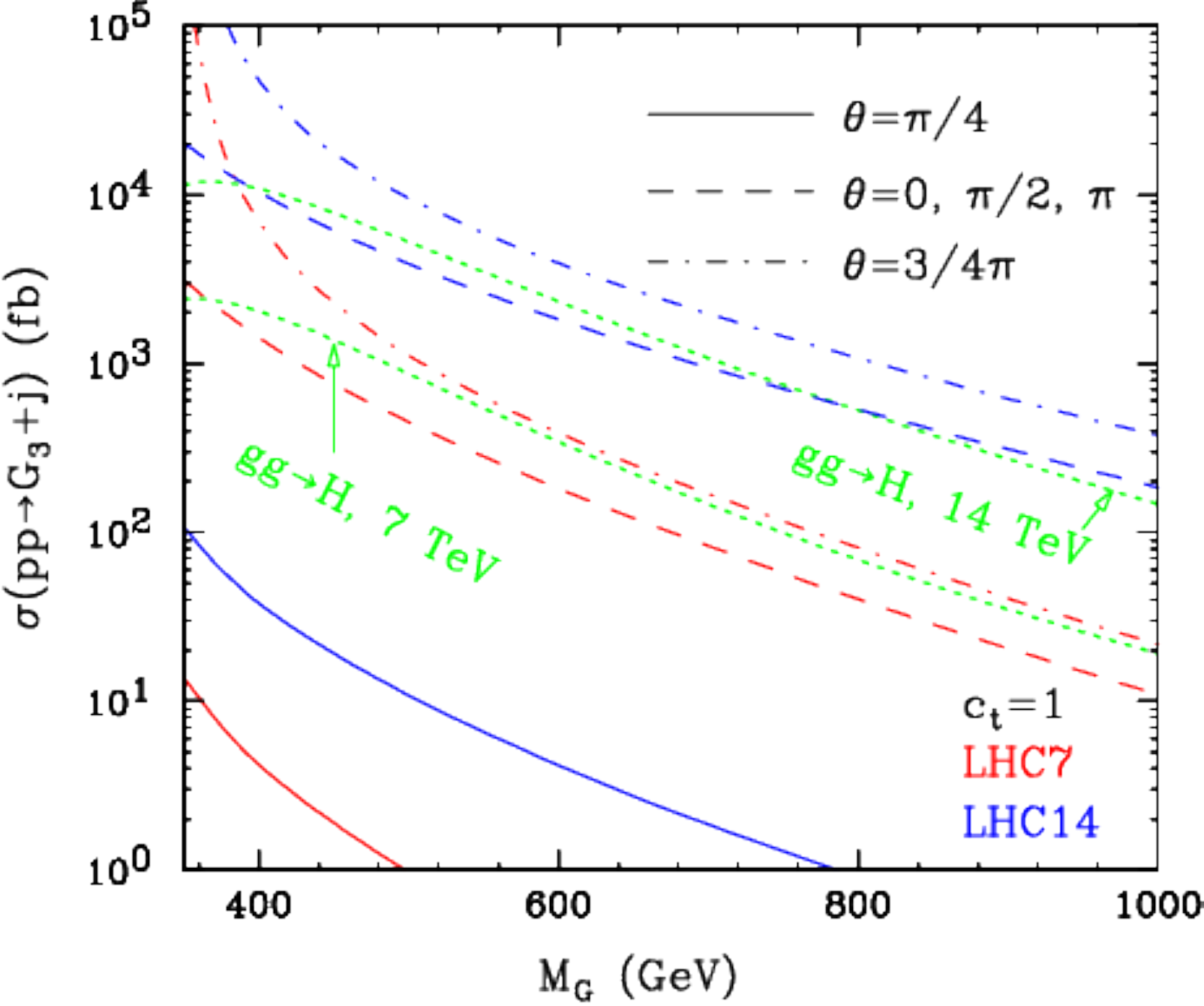}
  \caption{\label{fig:xsecG1j}
    Cross section of the loop induced production mode $pp\to G_3+j$
    for various values of $\theta$ and utilizing $c_t=1$. The jet
    defining requirements are given in Eq.~\eqref{eq:setup}. The red
    and blue curves depict the results for an LHC center-of-mass
    energy of $7\tev$ and $14\tev$, respectively. For comparison, the
    Higgs boson production cross section in gluon fusion with
    $M_H=M_G$ is also shown for both collision energies (cf.~the
    green dotted lines). 
  }
\end{figure}

Two different subprocesses contribute, namely
\begin{eqnarray}
 gg      &\to& G_3\,(\to t\bar t) + g~,\\
 q\bar q &\to& G_3\,(\to t\bar t) + g~,
\end{eqnarray}
and the remaining subprocesses can be obtained by crossing. For the
gluon channel ($gg$), we find 24 diagrams contributing; for the quark
channel ($q\bar q$), there are only two. The topology for the
quark--gluon channel ($qg$ or $\bar qg$) is the same as in the
$q\bar q$ case. Note that in the quark channel, no initial state
radiation can occur as these contributions are zero due to vanishing
color factors. A sample of gluonic diagrams and the two individual
quark initiated diagrams are depicted in Figure~\ref{diagsttj}.

In the quark initiated subprocess the only contribution is via an odd
number of particles attached to the top quark loop. In case all three
particles are vector-like particles, this contribution vanishes due to
Furry's theorem. This is the case if $\theta=\pi/4$, as for this case
there is no axial component in the $G_3$ resonance. We checked this
behavior numerically, which is an important consistency check for our
calculation. In the gluon initiated case the contribution does not
completely vanish for $\theta=\pi/4$ as there are still contributions
from box diagrams.

Our results are shown in Figure~\ref{fig:xsecG1j} at collider energies
of $7\tev$ and $14\tev$ for $c_t=1$. Avoiding the chirality suppressed
regime ($\theta=\pi/4$), these cross sections are comparable to the
Higgs boson production cross section over a wide range of $M_G$. At
$14\tev$, they can even become a few times larger. The strong
dependence on chirality is noticed especially at $\theta=\pi/4$, for
reasons as explained above.

\boldmath
\subsubsection{Off-shell production of the $G_3$ resonance:
  $pp\to G_3\to t\bar t$}
\unboldmath\label{offshell}

\begin{figure}[t!]
  \centering
  \includegraphics[width=0.8\textwidth]{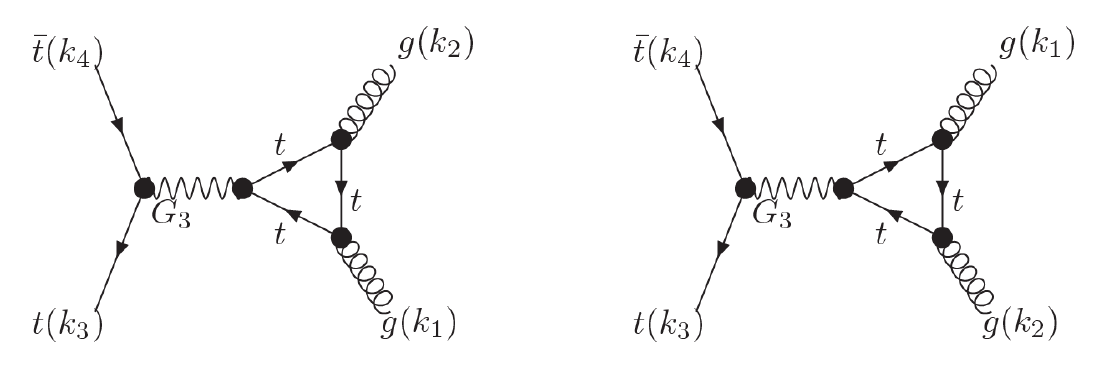}
  \caption{\label{diagstt}
    The two diagrams that contribute, at lowest order, to the loop
    induced production of $G_3\,(\to t\bar t)$ plus no extra parton.
    The quark initiated subprocess vanishes due to the color
    algebra.}
\end{figure}

In the case of loop induced, exclusive $G_3$ production, all
contributions vanish as long as the $G_3$ is produced on-shell. This
is a consequence of the Landau--Yang theorem. However, there is a
non-negligible contribution from the $G_3$ being off-shell which gives
a non-zero result. Including the $G_3$ decay into $t\bar t$, we only
find two diagrams, which are depicted in Figure~\ref{diagstt}. They
each contribute to the gluon initiated subprocess of the $t\bar t+0$
jet channel. No other partonic subprocess exists, and we verified
numerically that even the $gg$ initial state contribution turns zero
in the on-shell case. The latter moreover means that we passed another
important check of our calculation.

\begin{figure}[t!]
  \centering
  \begin{subfigure}[b]{0.49\textwidth}\centering
    \includegraphics[width=1\textwidth]{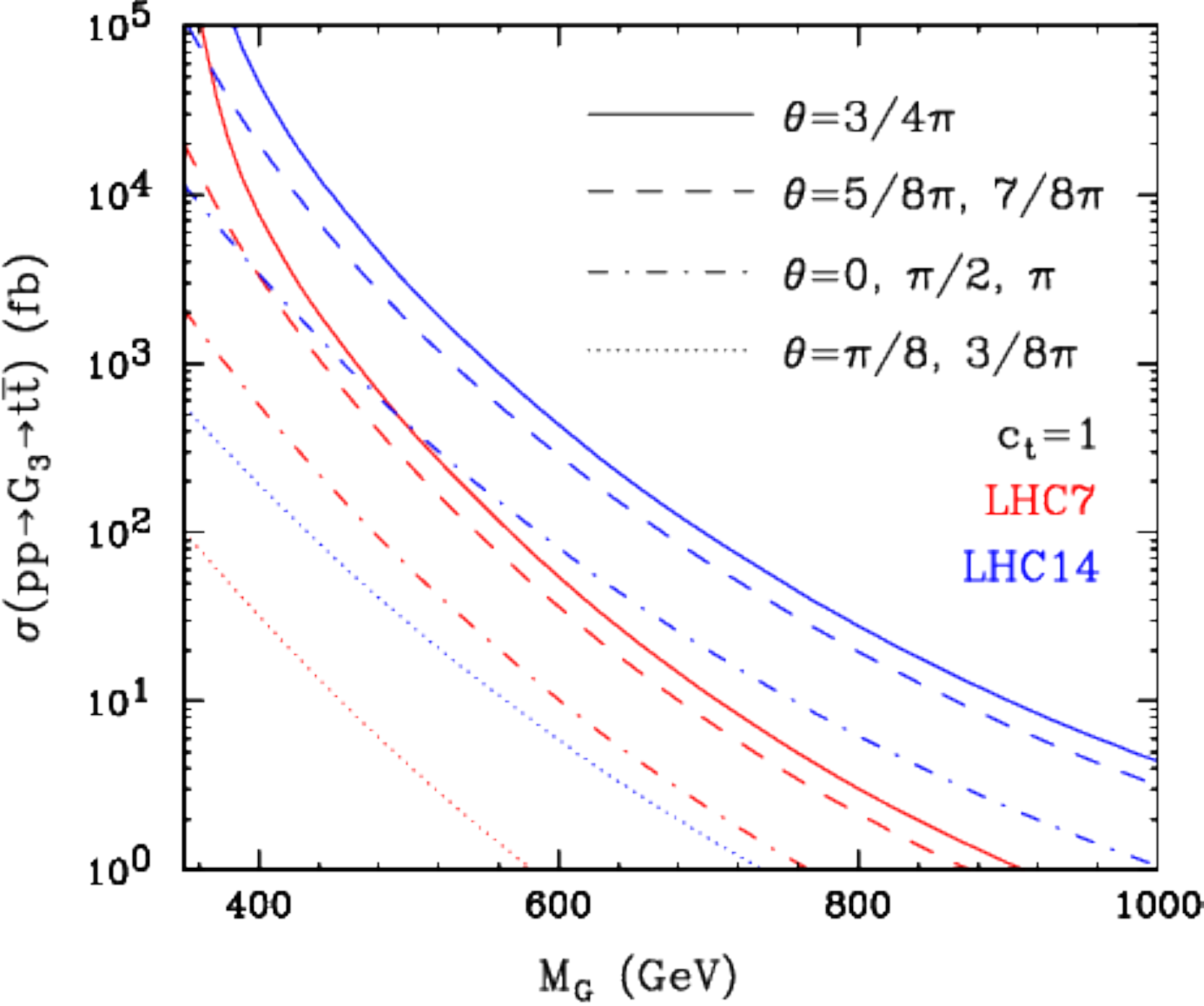}
    \caption{\label{sfig:xsecG0j}}
  \end{subfigure}
  \hskip1mm
  \begin{subfigure}[b]{0.49\textwidth}\centering
    \includegraphics[width=1\textwidth]{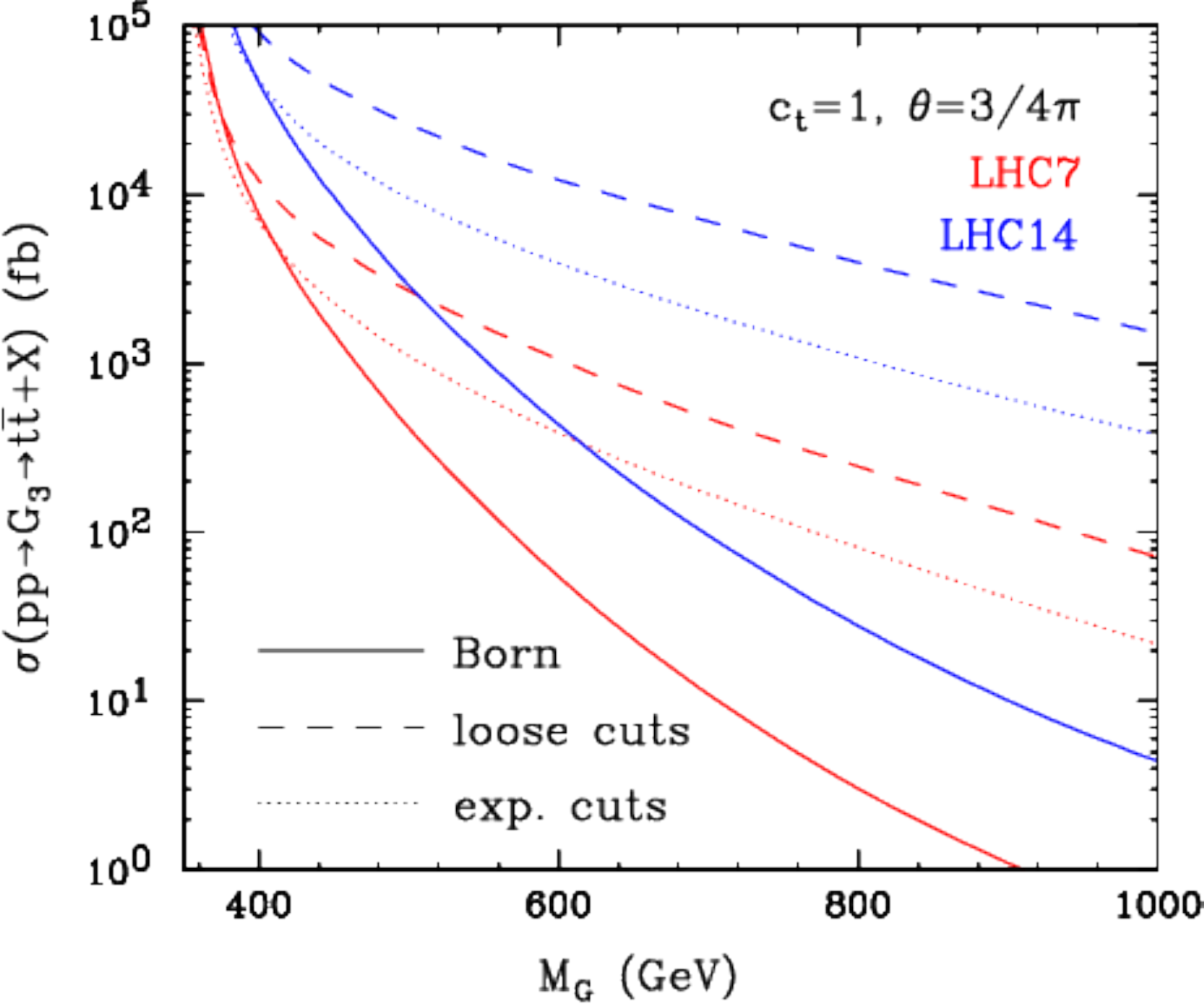}
    \caption{\label{sfig:xsecG0jcompG1j}}
  \end{subfigure}
  \caption{\label{fig:xsecG0j}
    Production cross section for (\ref{sfig:xsecG0j}) the loop induced
    process $pp\to G_3\to t\bar t$ (the Born process) using
    $c_t=1$ and various values of $\theta$. In the right panel
    (\ref{sfig:xsecG0jcompG1j}), the maximal cross sections, obtained
    by choosing $\theta=\frac{3}{4}\,\pi$, are compared to those for
    the loop induced process with an additional parton in the final
    state. While the ``experimental cuts'' are given in
    Eq.~\eqref{eq:setup}, the ``loose cuts'' are defined by
    $p_{T,j}>20\gev$ and a fairly wide jet rapidity window of
    $|\eta_j|<6.0$.}
\end{figure}

Figure~\ref{sfig:xsecG0j} can be directly compared to
Figure~\ref{fig:xsecG1j} as it uses the same layout to show the cross
section dependence on the mass of the $G_3$ particle for our choice of
$c_t=1$ and several different values of the chirality angle
$\theta$. As can be seen from these figures, one important consequence
of the model -- which at first seems unusual -- is the sizeable
increase of the cross section for associated jet production with
respect to that of the pure $G_3\to t\bar t$ process. Their
relative importance strongly depends on the mass of the $G_3$, and
varies from being roughly of the same size for $M_G=400\gev$ up to
about two orders of magnitude for $M_G=1\tev$. This is nicely
demonstrated in Figure~\ref{sfig:xsecG0jcompG1j} for the choice of
$\theta=3\pi/4$, which maximizes the cross section in both cases.

One reason for the enhancement of the $t\bar t+\mbox{jet}$ channel is
the appearance of box diagrams in the $t\bar t+1$ parton processes.
These box diagrams are not governed by Furry's theorem owing to the
even number of spin-1 particles attached to the loop. Another reason
is that initial state radiation can shift the gluon attached to a
triangle off-shell, thus enabling the $G_3$ to go on-shell. It is
therefore expected that the $G_3\,j$ associated production will
contribute substantially to the inclusive cross section determination
based on $t\bar t$\/ final states. Moreover, even for an exclusive
$t\bar t+0$ jet measurement, there will be a contribution stemming
from the case where the jet is not resolved, and this contribution
cannot be neglected in the zero-jet bin. Because of these facts, we
will have to combine the parton level calculations for $G_3+0$ and
$G_3+1$ extra parton, to arrive at an ``approximate NLO''
prediction. Only after we can use this prediction to apply the cross
section bounds from inclusive and exclusive $t\bar t+0$ jet
measurements to constrain the $G_3$ model parameters. We return to
this discussion in more detail in the next section.

\section{Current Experimental Bounds and Future Prospects for LHC Run\:2}
\label{bounds}

We will use results from searches in multi-top quark final states and
recent experimental measurements related to top quark pair production,
also in association with jets, to constrain the parameter space (mass
and couplings or chirality) for the production of the $G_3$ resonance.
Cross section limits are among the strongest handles that we can
utilize to extract current bounds from the Tevatron as well as from
the LHC for $7$~and~$8\tev$ collision energies. Taken these bounds, we
have to make sure to choose the model parameters such that our signal
evades all of these bounds; only after we can discuss the prospects of
the model we are considering here for resonance searches at the
$14\tev$ LHC.

If we compare the cross sections between the tree level and one-loop
production modes, we notice a clear ranking in favor of the loop
induced modes. Roughly speaking, they are separated by at least two
orders of magnitude for light $G_3$ resonances ($M_G\approx400\gev$)
where we obtain ${\cal O}(10\pb)$ cross sections in the loop induced
production channels. For the heavier $G_3$ (with $M_G\approx1\tev$),
we find the $G_3\,j$ one-loop mode dominating over all other channels
by a factor of about 50. The pure $G_3$ mode drops much faster and
becomes comparable to the production at the tree level, which is of the
order of $10\fb$ for LHC $14\tev$ but remains well below $1\fb$ for
LHC Run\:1 energies. The loop induced modes are therefore essential in
a search for top-philic resonances, especially for a test of the high
mass regime where these modes become indispensable. This does not mean
that we should discard the tree level production channels right away.
They give rise to more exotic final states characterized by multi-top
quark signatures that do not have to compete with the larger
backgrounds, which occur for loop level generated $(G_3\to)\,t\bar t+0,1$
jet final states. Because of their peculiar signatures, the tree level
modes carry valuable information and provide a great opportunity for
diversifying the search, in particular for low mass top-philic
resonances.

\subsection{Bounds from searches for multiple top quark final states}

All tree level production modes lead to either three top quarks
($G_3\,tj$ and $G_3\,tW$) or four top quarks ($G_3\,t\bar t$) in the
final state. This prominent feature of abundant top quarks needs to be
exploited in searching for distinct signs of decaying
$G_3$ resonances.

\begin{figure}[t!]
  \centering
  \includegraphics[width=0.67\textwidth]{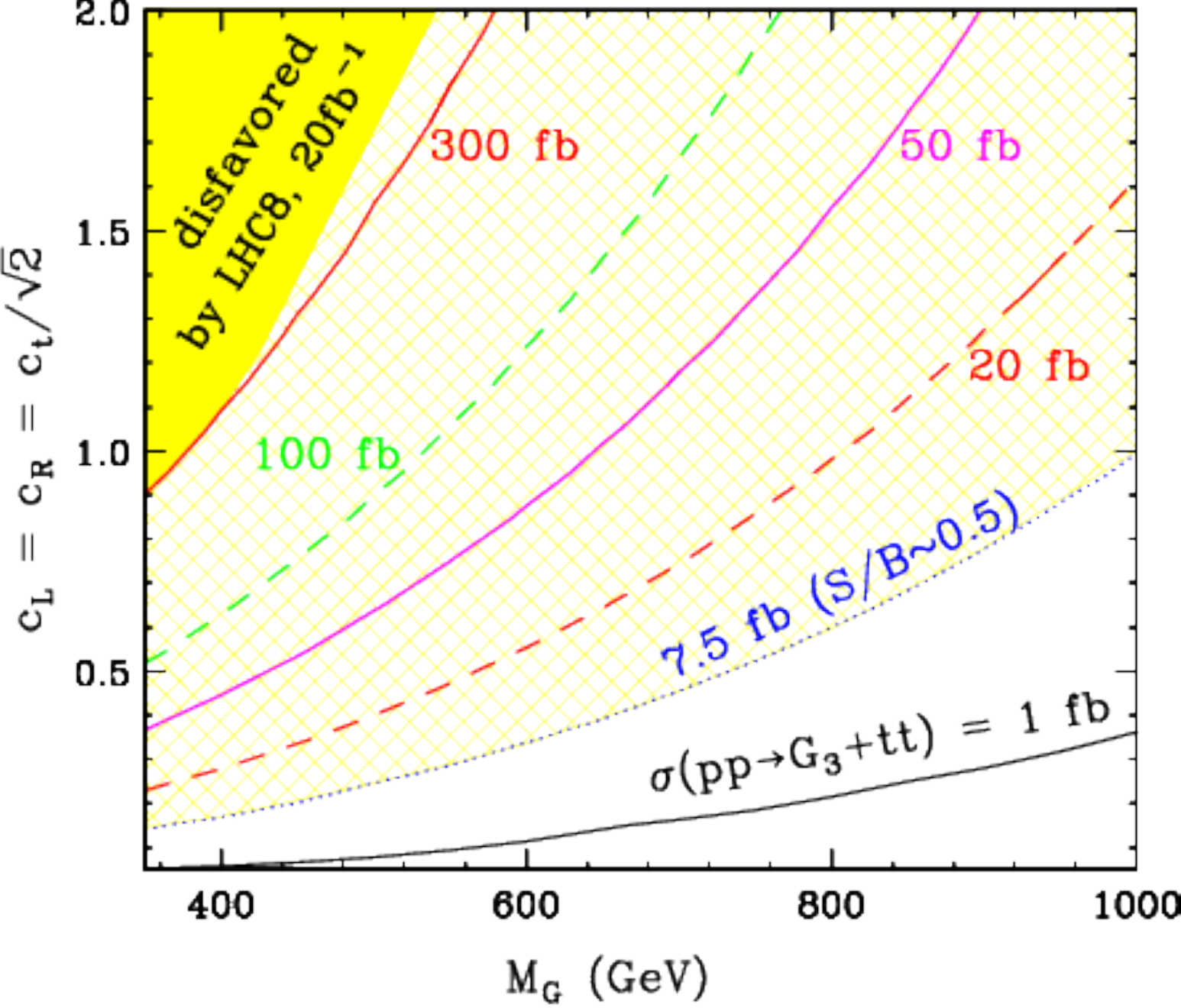}
  \caption{\label{fig:G3_bound}
    Production cross section at the $14\tev$ LHC for the tree level
    process $pp\to G_3+t\bar t$ shown in the model parameter plane of
    $M_G$-$c_t/\sqrt 2$. The upper left corner (shaded in yellow) of
    this parameter space has been already probed by the $8\tev$ LHC.
    The lighter shade is used to indicate an exclusion potential that
    may be achieved for Run\:2 of the LHC.
  }
\end{figure}

In particular the largest production channel at the tree level, the
$G_3\,t\bar t$ process results in a final state with four top quarks,
which is constrained by a CMS study~\cite{Khachatryan:2014sca}.%
\footnote{Similar studies have been carried out by ATLAS providing a
  weaker limit for the $7\tev$ LHC~\cite{ATLAS-CONF-2012-130}.}
The CMS upper limit on the four top-quark production cross section is
$32\fb$ (95\%~C.L.) at LHC $8\tev$ with $19.6\fbi$. This constraint is
depicted in the top left corner of Figure~\ref{fig:G3_bound} by the area
that has been shaded in plain yellow. The (other) contours in the
$M_G$-$c_L$ parameter plane where $c_L\equiv c_R=c_t/\sqrt 2$
represent the production cross section of four top quarks at an LHC
energy of $14\tev$.\footnote{As discussed in Section~\ref{tree}, the
  chirality ($\theta$) dependence of the cross section is very weak,
  calling for the obvious choice to show the cross section as a
  function of the model parameters $M_G$ and $c_t$.}
From this visualization, we note that most of the parameter space
remains unconstrained, yet becomes conveniently testable at Run\:2 for
rather natural choices of the coupling ($c_L\sim{\cal O}(1)\sim c_R$)
and $G_3$ masses below $\sim1000\gev$.

At $14\tev$ the SM prediction for four top quarks at NLO is around
$15\fb$ (see Ref.~\cite{Bevilacqua:2012em}). This means that the
contribution from the $G_3$ signal can lead to a significant
enhancement of the four top-quark cross section yielding a large
signal over background ratio ($S/B$) for the major fraction of the
parameter space. If we take the scale variation as a measure for the
systematic uncertainty of the NLO calculation, which the authors
of Ref.~\cite{Bevilacqua:2012em} determined to be around $\pm4\fb$,
the measurement in the four top-quark channel will unavoidably be
dominated by systematic uncertainties. A meaningful search thus
requires a large enough $S/B$ of around $0.5$. This yields a rough
estimate for the potential of a Run\:2 search, which we mark in
Figure~\ref{fig:G3_bound} using the lighter shade of yellow. Note that
the $S/B$ requirement is more stringent, as the statistical
significance ($S/\sqrt B$) for the discovery of a potential signal of
$7.5\fb$ can be easily reached with as little as $7\fbi$ of data.

\subsection{Bounds from uncertainties in top quark pair production measurements}

The one-loop production modes can be traced in a very general way
through their $t\bar t+\mbox{jets}$ signatures. In both cases
considered here, the $G_3$ vector particles decay into rather
energetic top quark pairs, which will likely be accompanied by one jet
owing to the dominance of the $G_3\,j$ production channel.
This is true in particular for heavy resonances above $M_G=600\gev$,
cf.~Figure~\ref{sfig:xsecG0jcompG1j}. It is of course no secret that
$t\bar t+\mbox{jets}$ final states are populated by a fairly
large number of background processes but from experimental
measurements designed to track down $t\bar t$ pairs, we already have a
great deal of experience to which kinematical requirements work best
in such an environment. In our case, we thus will be confronted with
SM top quark pair production as the by far largest contribution to the
portfolio of backgrounds. At zeroth order, we hence are in the position
to neglect the other processes such as $W+\mbox{jets}$, and just rely
on the excellent theoretical and experimental control of the $t\bar t$
production~\cite{Czakon:2013goa} to obtain reasonable estimates for
the detection of the imprints from decaying $G_3$ resonances.

Considering the production of $t\bar t$ final states, we have to bear
in mind that they may emerge in association with a number of $X$ jets,
denoted conveniently by $t\bar t+X$. In a fully inclusive measurement,
all jet contributions, $X=0,1,\ldots$, are taken into account while
for an exclusive $t\bar t+X$ analysis, one requires to find exactly
$X$ jets accompanying the top quark pair. In general there are three
different levels of experimental exclusiveness which are roughly
characterized by (1)~inclusive total $t\bar t$ cross section
measurements, (2)~differential measurements in the number of jets
associated with the $t\bar t$ production and (3)~measurements of
differential distributions based on pair and single top quark
properties such as the invariant mass or the transverse momentum.
Here, we will exploit the first two categories only, leaving us with
plenty of data from which we can infer important constraints on the
$G_3$ model. To distinguish them more easily in our discussion, we
introduce the following notation: ``(I)''~stands for the inclusive
measurements of point~(1) while ``(E)''~denotes the exclusive ones
under point~(2).

The third lever arm that we could utilize is given by differential
distributions based on $t\bar t$ system or single top quark objects.
Related observables can then be used to check on individual bins,
which should better be unaffected by signal contributions, at least
beyond a two-sigma deviation. However, these investigations are
outside the scope of this paper and their outcome will be discussed in
a forthcoming publication.

\subsubsection{Inclusive jet cross sections}

Measurements in the two lowest $t\bar t\ +\ge X$ jet bins are of
importance to our discussion, namely the production of $t\bar t$
final states plus any number of jets ($t\bar t\ +\ge0$ jets), which we
label ``(I0)'', and those occurring in association with at least one
jet ($t\bar t\ +\ge1$ jet), which we label ``(I1)''. Accordingly, we
can extract two types of limits from the respective uncertainties on
the cross section measurements: that is the inclusive zero-jet limit
and the inclusive one-jet limit, which we denote
$f\times\Delta\sigma^{(\ge0)}_{t\bar t}$ and
$f\times\Delta\sigma^{(\ge1)}_{t\bar t}$, respectively. The factor $f$
indicates that we have some leeway in applying these limits.
We certainly are on the safe side with factors of about two, allowing
the signal to grow no more than into a two-sigma deviation.

Focusing first on the entirely inclusive scenario (I0), we have to
confront the zero-jet limit with a prediction for the loop induced
production of $G_3$ that is fully accurate at NLO in QCD. In a
rigorous approach, this order of precision is forced upon us because
of the largeness of the one-jet contribution with respect to the
contribution obtained at the Born level. This peculiarity of our
signal model has been demonstrated in Figure~\ref{sfig:xsecG0jcompG1j}
but it may also be understood as a strong motivation to calculate the
two-loop corrections to our $G_3$ signal (for which we use the
notation ``$S\,$'' here). A two-loop accurate treatment however is
not within the scope of this paper and we have to resort to an
alternative. Knowing that a signal cross section description via
\beq
\sigma^{(\ge0)}_{S,\,\srm{LO}}\;=\;\sigma_\mrm{Born}
\qquad\mbox{or}\qquad
\sigma^{(\ge0)}_{S,\,\srm{NLO}}\;=\;\sigma_\mrm{Born}+
  \sigma_\mrm{Virt}+\sigma_\mrm{Real}
\eeq
cannot be trusted or remains, respectively, out of reach for now, we
sign up for a compromise between the two, which we will refer to as an
approximate NLO treatment of the signal.

We can write the signal cross section contributing to the inclusive
zero-jet bin as
\begin{equation}\label{eq:inc0}
\sigma^{(\ge0)}_{S,\,\srm{NLO}}\;=\;
\sigma_\mrm{Born}+\sigma^\mrm{loose.cuts}_\mrm{Real}+\delta\;\le\;
f\times\Delta\sigma^{(\ge0)}_{t\bar t}
\end{equation}
where we use
\begin{equation}
\delta\;=\;\sigma_\mrm{Virt}+\sigma^\mrm{IR}_\mrm{Real}
\end{equation}
to collect all terms, which cannot be calculated within the scope of
this paper. Note that Eq.~\eqref{eq:inc0} already shows the bound to
the signal inclusive $t\bar t$ cross section. To motivate, moreover
justify our simplified approach, it is useful to consider the
different regions of the $p_{T,t\bar t}$ distribution. The Born term
which we calculate from the $t\bar t+0$ parton amplitudes (at LO)
contributes to the $p_{T,t\bar t}=0$ bin, which also receives virtual
corrections as accounted for by $\sigma_\mrm{Virt}$ occurring in the
$\delta$ term. The real-correction term in Eq.~\eqref{eq:inc0} again
can be calculated (at LO), this time from the $t\bar t+1$ parton
amplitudes. The label ``$\mrm{loose.cuts}$'' is chosen to express a
careful separation between the fixed-order and resummation
$p_{T,t\bar t}$ regime that can be obtained by applying rather loose
jet cuts, which are more inclusive than common experimental jet
definitions, yet safe enough to ensure a meaningful LO description of
the $p_{T,t\bar t}$ spectrum above the Sudakov region (i.e.~for
$p_{T,t\bar t}\gtrsim20\gev$). The Sudakov region itself cannot be
treated without the proper inclusion of resummation effects below
$p_{T,t\bar t}\approx20\gev$. The soft(-$p_{T,t\bar t}$)
real-correction term, $\sigma^\mrm{IR}_\mrm{Real}$, is therefore
part of the $\delta$ correction to $\sigma^{(\ge0)}_{S,\,\srm{NLO}}$.
Of course, $\delta$ is separately finite by construction as it
contains (and cancels out) all IR divergences arising from the virtual
as well as the real corrections.

It helps our discussion to re-order Eq.~\eqref{eq:inc0} as follows:
\begin{equation}\label{eq:inc0estimate}
\left(\frac{\sigma_\mrm{Born}+\sigma^\mrm{loose.cuts}_\mrm{Real}}
     {\Delta\sigma^{(\ge0)}_{t\bar t}}\right)\times\,
\left(1+\frac{\delta}{\sigma_\mrm{Born}+\sigma^\mrm{loose.cuts}_\mrm{Real}}
\right)\;\le\;f~.
\end{equation}
The first bracket on the left-hand side is now fully controlled by our
parton level calculations, and we can individually constrain it, for
example, to be smaller than two. For the second bracket, it is not
unreasonable to assume that its cross section ratio varies no more
than $\pm50\%$, or very conservatively $\pm100\%$. This leaves us with
an estimate for the left-hand side of Eq.~\eqref{eq:inc0estimate}
reading $2\times(1\pm0.5)=1,\ldots,3\le f$, which is completely within
the cross section range that cannot be excluded owing to the
measurement uncertainties. Thus, although we are forced to speculate
by and large about the size of the (unknown) $\delta$ contribution, we
conclude that in the realm of (I0) analyses it is sufficiently safe to
set bounds according to
\begin{equation}\label{eq:inc0limit}
\sigma_\mrm{Born}+\sigma^\mrm{loose.cuts}_\mrm{Real}\;\le\;
f\times\Delta\sigma^{(\ge0)}_{t\bar t}~.
\end{equation}
This means, instead of constraining the Born cross section only, the
inclusive zero-jet limit has to be used to constrain the summed up
cross section for both the $t\bar t+0$ and $t\bar t+1$ parton
calculations. The looser phase-space definition of the jet in the
one-parton process enables us to include some of the soft
$p_{T,t\bar t}$ contributions that usually remain unresolvable in a
realistic jet analysis, even though we are clearly not in the position
to provide a full NLO treatment of the signal.

Turning to the discussion of the inclusive one-jet limits (I1), we
immediately recognize that an approximate NLO treatment for $G_3\,j$
would require us to compute the double-real corrections to the loop
level generation of $G_3$. As for the two-loop case, we leave this
calculation to a future study; also because we only expect moderate
real corrections to the $G_3\,j$ rate, which should be much smaller
than the drastic corrections that we found for the $G_3$ rate. This
assumption should be reasonable since there is no mechanism in place,
based on Furry's theorem, that could suppress the Born level of the
$G_3\,j$ process in a way comparable to the suppression of the $G_3$
Born process. Based on these arguments, it is absolutely sufficient to
directly apply the one-jet limit to the $t\bar t+1$ parton (LO)
calculations obeying (as closely as possible) the jet requirements as
given by the particular experiment. Hence, we can summarize this
situation as
\begin{equation}\label{eq:inc1limit}
\sigma^{(\ge1)}_{S,\,\srm{LO}}\;=\;
\sigma^\mrm{exp.cuts}_\mrm{Real}\;\le\;f\times\Delta\sigma^{(\ge1)}_{t\bar t}~,
\end{equation}
noting that
$\sigma^\mrm{exp.cuts}_\mrm{Real}<\sigma^\mrm{loose.cuts}_\mrm{Real}$
by construction -- in other words, the use of experimental cuts pushes
us further out in $p_{T,t\bar t}$, i.e.~further away from the Sudakov
region in the $p_{T,t\bar t}$ distribution as compared to the zero-jet
case.

\boldmath
\subsubsection{Exclusive jet cross sections and the $N_\mathrm{jets}$ distribution}
\unboldmath

In contrast to the previously discussed situation, differential
distributions in the number of jets, in short $N_\mathrm{jets}$
distributions, give us the opportunity to extract cross section limits
for exclusive $t\bar t+X$ final states. We denote these quantities by
$\Delta\sigma^{(X)}_{t\bar t}$ as opposed to $\Delta\sigma^{(\ge X)}_{t\bar t}$,
which are used to express the cross section uncertainties of the
inclusive jet bins. Again, the measurements of interest to us are
those targeting the $t\bar t+0$ jet bin (no extra-jet bin) and the
$t\bar t+1$ jet bin (exactly one-jet bin), which help us obtain
estimates for the exclusive zero-jet and one-jet limits,
$\Delta\sigma^{(0)}_{t\bar t}$ and $\Delta\sigma^{(1)}_{t\bar t}$,
respectively. Within our labelling scheme, we distinguish these two
cases by ``(E0)'' and ``(E1)''.

Based on our findings for the (I0) case, we can easily derive an NLO
approximate relation that obeys the kinematical requirements on (E0)
final states. To incorporate the zero-jet exclusiveness, we modify
Eq.~\eqref{eq:inc0limit} by removing the resolved-jet contribution
from the rate given by $\sigma^\mrm{loose.cuts}_\mrm{Real}$. At LO,
this simply means that we subtract the $t\bar t+1$ parton cross
section for experimental jet cuts from the left-hand side of
Eq.~\eqref{eq:inc0limit}. In the (E0) category, we therefore only
constrain the leftover part, i.e.~the unresolved part of the
respective (I0) relation, and hence arrive at
\begin{equation}\label{eq:exc0limit}
  \sigma_\mrm{Born}+\sigma^\mrm{loose.cuts}_\mrm{Real}-
  \sigma^\mrm{exp.cuts}_\mrm{Real}\;\le\;
  f\times\Delta\sigma^{(0)}_{t\bar t}~.
\end{equation}
Note that there are no changes to the loose cut definition; the loose
jet constraints are chosen as before to keep the resummation effects
negligible.

Similarly to (I1), any exclusive one-jet limit can be used to place
bounds on the $t\bar t$ plus one-parton cross section. Already in the
context of a leading-order calculation, this is a reasonable approach
for the same reasons as previously described for the case of (I1). All
we need to constrain is the resolved contribution of
$\sigma^\mrm{loose.cuts}_\mrm{Real}$, and practically there is no
difference whether we consider (E1) or (I1) as long as the $t\bar t+1$
parton calculation has no more than LO accuracy.
Equation~\eqref{eq:inc1limit} therefore shows almost no difference to
what we write down for the case of (E1) where we have
\begin{equation}\label{eq:exc1limit}
  \sigma^\mrm{exp.cuts}_\mrm{Real}\;\le\;f\times\Delta\sigma^{(1)}_{t\bar t}~.
\end{equation}
The difference between this LO expression and Eq.~\eqref{eq:inc1limit}
originates merely from the specifics of the (experimental) jet
definition applied in both the exclusive and inclusive scenarios.
These kinematical choices determine whether the extra parton gets
``resolved'' as a jet or missed in the bulk of soft/collinear
radiation.

Exclusive jet cross sections are rarely reported directly. We have to
infer them from $N_\mrm{jets}$ distributions, usually presented in
terms of exclusive $X$-jet fractions for $\ell+\mbox{jets}$ and dilepton
final states confined to a fiducial volume. In some cases, these jet
fractions, $r_X$, are also provided for the distribution of additional
jets occurring in $t\bar t+X$ production. The computation of an
exclusive $X$-jet cross section is thus expressed as
\begin{equation}
  \sigma^{(X)}_{t\bar t}\;=\;r_X\times\sigma^{(\ge0)}_{t\bar t}~,
\end{equation}
and requires knowledge of $\sigma^{(\ge0)}_{t\bar t}$, the total
inclusive rate for $t\bar t$ production, which can be taken from
flagship measurements that release this cross section after the
application of unfolding and acceptance corrections. The uncertainty
associated with the exclusive jet cross sections is hence given as
\begin{equation}\label{eq:rX2lim}
  \Delta\sigma^{(X)}_{t\bar t}\;=\;
  \left(\frac{\Delta r_X}{r_X}+
  \frac{\Delta\sigma^{(\ge0)}_{t\bar t}}{\sigma^{(\ge0)}_{t\bar t}}\right)
  \,\times\;r_X\times\sigma^{(\ge0)}_{t\bar t}~,
\end{equation}
which we have to take into account when we determine the cross section
limits in the next subsection. Note that Eq.~\eqref{eq:rX2lim} is
written for the more conservative approach of summing the two
individual uncertainties linearly rather than adding them in
quadrature as one does for uncorrelated Gaussian errors.

\subsubsection{Constraints on the loop level resonance production}

Given the four types of limits discussed above, we worked out how to
apply them to our specific calculations. We now need to quantify the
exact size of the inclusive and exclusive cross section uncertainties
$\Delta\sigma^{(\ge X)}_{t\bar t}$ and $\Delta\sigma^{(X)}_{t\bar t}$,
respectively, using current experimental results.
Table~\ref{tab:limits} lists the values, which we determined in order
to make the relations \eqref{eq:inc0limit}, \eqref{eq:inc1limit},
\eqref{eq:exc0limit} and \eqref{eq:exc1limit} explicit. As explained
before, we argued for $f=2$ as a safe choice in fixing the $f$-factor.
Our results however will be reported for the more restrictive case of
using $f=1$.

\begin{table}[t!]
  \centering\small
  \begin{tabular}{lrccccccc}\toprule\\[-8pt]
    Collider & Category:  & I0    && I1   && E0         && E1 \\\midrule\\[-8pt]
    Tevatron $1.96\tev$  && $0.41$&&$0.54$&&$0.81$            \\[2pt]
    LHC $7\tev$          && $10$  &&$26$  &&$8,\ldots,9$&&$12$\\[2pt]
    LHC $8\tev$          && $13$ \\\bottomrule
  \end{tabular}
  \caption{\label{tab:limits}%
    Overview of the $\Delta\sigma_{t\bar t}$ quantities, in pb, for
    the different cross-section limit categories concerning current
    inclusive (I$X$) and exclusive (E$X$) cross section measurements
    for $t\bar t$ production in association with $X=0,1$ jets.}
\end{table}

The Tevatron limits in Table~\ref{tab:limits} have been obtained, for
(I0), from the combination of measurements with the goal to determine
the top quark pair production cross section at $1.96\tev$ and, for
(I1) and (E0), from a CDF measurement of the $t\bar t+\mbox{jet}$
cross section with $4.1\fbi$ of Tevatron data. The results and their
related uncertainties have been reported in
Refs.~\cite{Aaltonen:2013wca} and \cite{CDF-NOTE-9850}, respectively.
We have checked that these limits are of no consequence for the
hadronic production of the $G_3$ as the gluon initiated subprocesses
cannot be tightly constrained at the Tevatron. As a matter of fact, to
cross (or reach) the production threshold of $t\bar t$ pairs, the
Tevatron was forced into the operational mode of a $q\bar q$ collider,
leaving obviously little room to test the highly important gluon
production channels of the $G_3$ model. In contrast to the Tevatron,
the LHC predominantly operates as a $gg$ collider. It is the more
natural place to look for $G_3$ resonances and therefore enables us to
set stronger limits on ($G_3$ induced) deviations from SM $t\bar t$
production. Thus, all other cross section limits shown in
Table~\ref{tab:limits} have been extracted from a variety of LHC
measurements.

Considering the most inclusive, the (I0) case first, we have several
comparable cross section measurements from the ATLAS and CMS
collaborations for both Run\:1 energies, in the single lepton plus
jets ($\ell+\mbox{jets}$) channel as well as the dilepton ($\ell\ell$)
channel~\cite{ATLAS-CONF-2012-131,Chatrchyan:2012bra,Chatrchyan:2012ria,%
  Chatrchyan:2013faa,Aad:2014kva,Aad:2014jra,ATLAS-CONF-2012-149,%
  CMS-PAS-TOP-12-006}.
For the $7\tev$ LHC, we find the CMS dilepton
measurement~\cite{Chatrchyan:2012bra} based on ${\cal L}=2.3\fbi$ to
be very accurate, but as a result of the large spread among the
different central values of all measurements, we have decided for a
reasonably safe compromise, which is to use the $10\pb$ uncertainty of
the LHC combined result published in September
2012~\cite{ATLAS-CONF-2012-134}.
For the $8\tev$ LHC (and an $m_t$ reference value of $172.5\gev$), the
top quark pair production cross section has been determined very
recently as $\sigma_{t\bar t}=(241.5\pm8.5)\pb$, with an additional
uncertainty of $4.2\pb$ owing to LHC beam effects~\cite{ATLAS-CONF-2014-054}.
This is the result of a first combination of ATLAS and CMS
measurements with exactly one electron and one muon in the final
state, which benefits from the fact that both the ATLAS and CMS
dilepton channels have delivered very precise results with
uncertainties marginally larger than
$10\pb$~\cite{ATLAS-CONF-2013-097,Aad:2014kva,Chatrchyan:2013faa}.
Again, the spread in all $8\tev$ measurements is at least of similar
size, and hence we have chosen, as before, to be more conservative in
our table where we included the somewhat larger uncertainty ($13\pb$)
of the CMS dilepton channel obtained after analyzing $5.3\fbi$ of
data~\cite{Chatrchyan:2013faa}.

From a theoretical point of view, it is interesting to note that the
(I0) numbers cannot be controlled more precisely than they have been
actually measured in the experiments. The current situation is
described adequately by the fact that experimental and theoretical
uncertainties are of the same size. This becomes clear by comparing
the (I0) limits in Table~\ref{tab:limits} with the total theory errors
($\delta_\mrm{tot}$) on $\sigma_{t\bar t}$ as published in
Refs.~\cite{Czakon:2013tha,Czakon:2013goa} for Tevatron Run\:II and
LHC Run\:1 energies.

For the other jet categories, (I1), (E0) and (E1), it is more
intricate to find suitable results, which work on the level of top
quarks and additional jets. This level of information requires several
steps of data corrections such as the unfolding of detector effects,
the kinematic reconstruction of top quarks and the extrapolation from
the acceptance region to the full phase space. These experimental
procedures are sophisticated, costly, and take a lot of work. Still,
there are a number of publications (and more will become available
with time), in particular for the $7\tev$ LHC, that help us specify
the limits for the categories other than (I0).

In Ref.~\cite{ATLAS:2012ooa}, the cross section for the production of
$t\bar t$ pairs in association with at least one jet has been reported
for the $7\tev$ LHC. The measurement results are given in a fiducial
volume as well as in the inclusive one-jet phase space defined by the
anti-$k_T$ jet algorithm. The latter, i.e.~the acceptance-corrected
result is used to assign a value to $\Delta\sigma^{(\ge1)}_{t\bar t}$
as shown in Table~\ref{tab:limits}. Concerning our calculations, we
have then made sure that the jet parameters of Eq.~\eqref{eq:setup}
like $p_{T,j}$, $\eta_j$ and $R$ exactly match those used by the
experiment.

The (E0) and (E1) $7\tev$ limits have been obtained from the results
given in section~7 of the CMS publication on jet multiplicity
distributions~\cite{Chatrchyan:2014gma}. Their jet finding procedure
is again based on the anti-$k_T$ algorithm, this time with slightly
different parameter settings reading $p_{T,j}>30\gev$, $|\eta_j|<2.4$
and $R=0.5$. The $N_\mrm{jets}$ distribution however is also provided
in terms of additional jets on top of the already reconstructed
$t\bar t$ system.\footnote{The corresponding ATLAS
  publication~\cite{Aad:2014iaa} as well as its earlier conference
  note~\cite{ATLAS-CONF-2012-155} do not contain $N_\mrm{jets}$
  information at the level of reconstructed top quarks and jets. In
  comparison to Ref.~\cite{Chatrchyan:2014gma} (cf.~Table~3 therein),
  we nevertheless use the $N_\mrm{jets}$ distributions provided by ATLAS
  in the $\ell+\mbox{jets}$ channel to verify that both collaborations
  indeed find similar sizes in the error accounting for the lower jet
  bins. However, we do not expect improvements to the limits derived
  from the CMS measurement as the uncertainties stated by ATLAS turn
  out to be somewhat larger.}
This is a big advantage for us as it avoids any discussion concerning
the top quark decay effects that we cannot simulate here for the $G_3$
signal. Without this reconstruction, the $G_3$ production particularly
affects the $2+3$-jet and the $4+5$-jet bins of the $\ell\ell$ and the
$\ell+\mbox{jets}$ channel, respectively, however in a complicated way
that is hard to approximate without the proper implementation of the
decays. As we learn from the results of Ref.~\cite{Chatrchyan:2014gma}
(cf.~Figure~4 therein), using an additional jet counting, little of
the zero-jet contribution is found in the $5$-jet bin of the
$\ell+\mbox{jets}$ channel where the decay jets are included to the jet
counting. The $t\bar t+1$ jet contribution however spreads rather
equally over the $4$-jet and $5$-jet bin. It is not clear a priori
whether the $G_3$ signal including its top quark decays will leave an
imprint similar to that given by ordinary $t\bar t$ production. This
emphasizes the importance of the CMS $t\bar t+\mbox{jets}$ data (given
in Table~4 and Figure~6) of Ref.~\cite{Chatrchyan:2014gma}. For the
exclusive jet fractions, $r_X$, we find $r_0=0.332$ and $r_1=0.436$
quoted with an uncertainty of $9.0\%$ and $9.8\%$, respectively. Based
on Eq.~\eqref{eq:rX2lim}, we then obtain the upper limits on the
signal exclusive-jet cross sections: $\Delta\sigma^{(0)}_{t\bar t}=8.5\pb$
and $\Delta\sigma^{(1)}_{t\bar t}=11.8\pb$ where we used
$\sigma^{(\ge0)}_{t\bar t}=(173\pm10)\pb$ in accordance with our
choice for the (I0) case.\footnote{The more aggressive,
  i.e.~quadrature summed version of Eq.~\eqref{eq:rX2lim} would yield
  slightly smaller exclusive jet cross section limits:
  $\Delta\sigma^{(0)}_{t\bar t}=6.1\pb$ and
  $\Delta\sigma^{(1)}_{t\bar t}=8.6\pb$.}
There is one caveat though; the results stated by CMS are strictly
valid only for a {\em visible}\/ phase-space definition (as clearly
described in the beginning of Section~6 of
Ref.~\cite{Chatrchyan:2014gma}). This visible phase-space definition
includes certain kinematical requirements on the top quark decay
products that we cannot implement here. In a first approximation, it
is however not unrealistic to assume that the CMS results carry over
to the {\em full}\/ phase space without taking any corrections into
account.

\begin{figure}[t!]
  \centering
  \begin{subfigure}[b]{0.49\textwidth}\centering
    \includegraphics[width=1\textwidth]{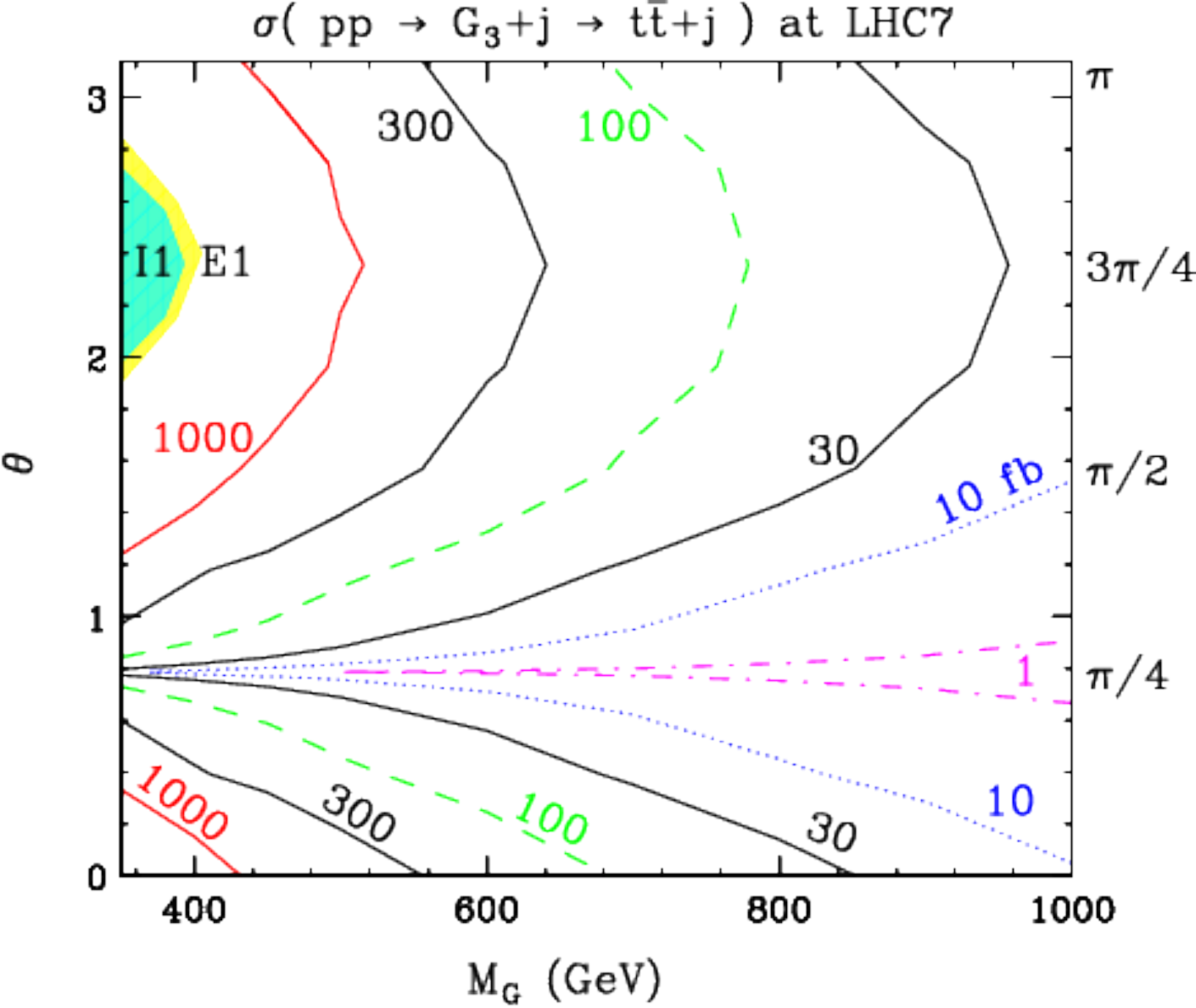}
    \caption{\label{sfig:G31j7tev}}
  \end{subfigure}
  \hskip1mm
  \begin{subfigure}[b]{0.49\textwidth}\centering
    \includegraphics[width=1\textwidth]{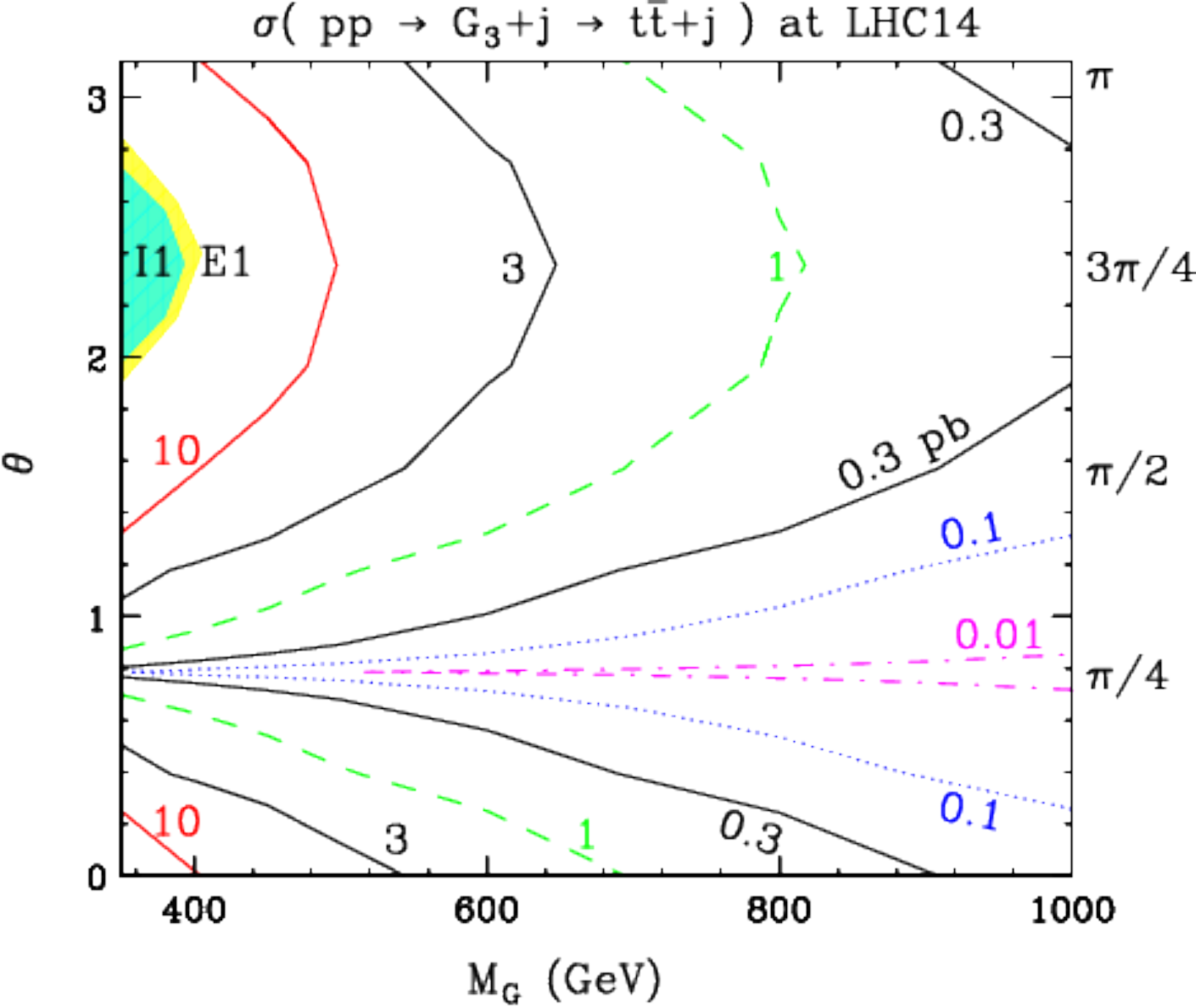}
    \caption{\label{sfig:G31j14tev}}
  \end{subfigure}
  \caption{\label{fig:G31j}
    Lines of constant production cross section in the $M_G$-$\theta$
    parameter plane using $c_t=1$ for the loop induced processes
    $pp\to G_3+j\to t\bar t+j$ at proton--proton colliders of
    $7\tev$~(\ref{sfig:G31j7tev}) and $14\tev$~(\ref{sfig:G31j14tev})
    center-of-mass energy. Results have been obtained employing the
    experimental jet cuts as summarized in Eq.~\eqref{eq:setup}. The
    excluded parameter space, at the one-sigma level, is indicated by
    the shaded areas (using the labels of Table~\ref{tab:limits}),
    with more explanations given in the text.}
\end{figure}

For the $8\tev$ LHC, we did find some useful information in
Ref.~\cite{CMS-PAS-TOP-12-041}, which however is not quite sufficient
to evaluate reasonable constraints that apply to the more exclusive
jet bin selections (I1), (E0) and (E1). We would need to make a number
of assumptions, in particular regarding acceptance corrections and the
kinematic effect of top quark decays, which eventually turns the whole
procedure of determining limits into a more or less speculative
exercise. To our knowledge, direct results of the type ``top quarks plus
jets'' at $8\tev$ are in the pipeline; however they have not been
published yet. It furthermore remains an open question whether such
new $8\tev$ limits could be immediately competitive with those
extracted from the $7\tev$ LHC measurements.

\bigskip
We start the discussion on parameter constraints with the $t\bar t+1\,j$
case. Figure~\ref{fig:G31j} shows the lines of constant cross sections
in a plane spanned by the model parameters $M_G$ and~$\theta$. The
left plot shows the results for $E_\trm{cm}=7\tev$ whereas the plot on
the right-hand side contains the predictions for $E_\trm{cm}=14\tev$.
Using Eq.~\eqref{eq:inc1limit} with $f=1$, we indicate cross sections
that are excluded at $7\tev$ at the one-sigma level. The green shaded
area shows the region in parameter space which is excluded by an
inclusive one-jet measurement (see (I1) in Table~\ref{tab:limits}). 
The more constraining yellow area displays the limit from exclusive
one-jet measurements, according to (E1). While in the $7\tev$ plot the
border of the shaded area corresponds to a line of constant cross
section, this does not have to be the case for the $14\tev$ plot. Here
the shaded area only marks a region in parameter space that is already
excluded by the $7\tev$ measurement.

\begin{figure}[t!]
  \centering
  \begin{subfigure}[b]{0.49\textwidth}\centering
    \includegraphics[width=1\textwidth]{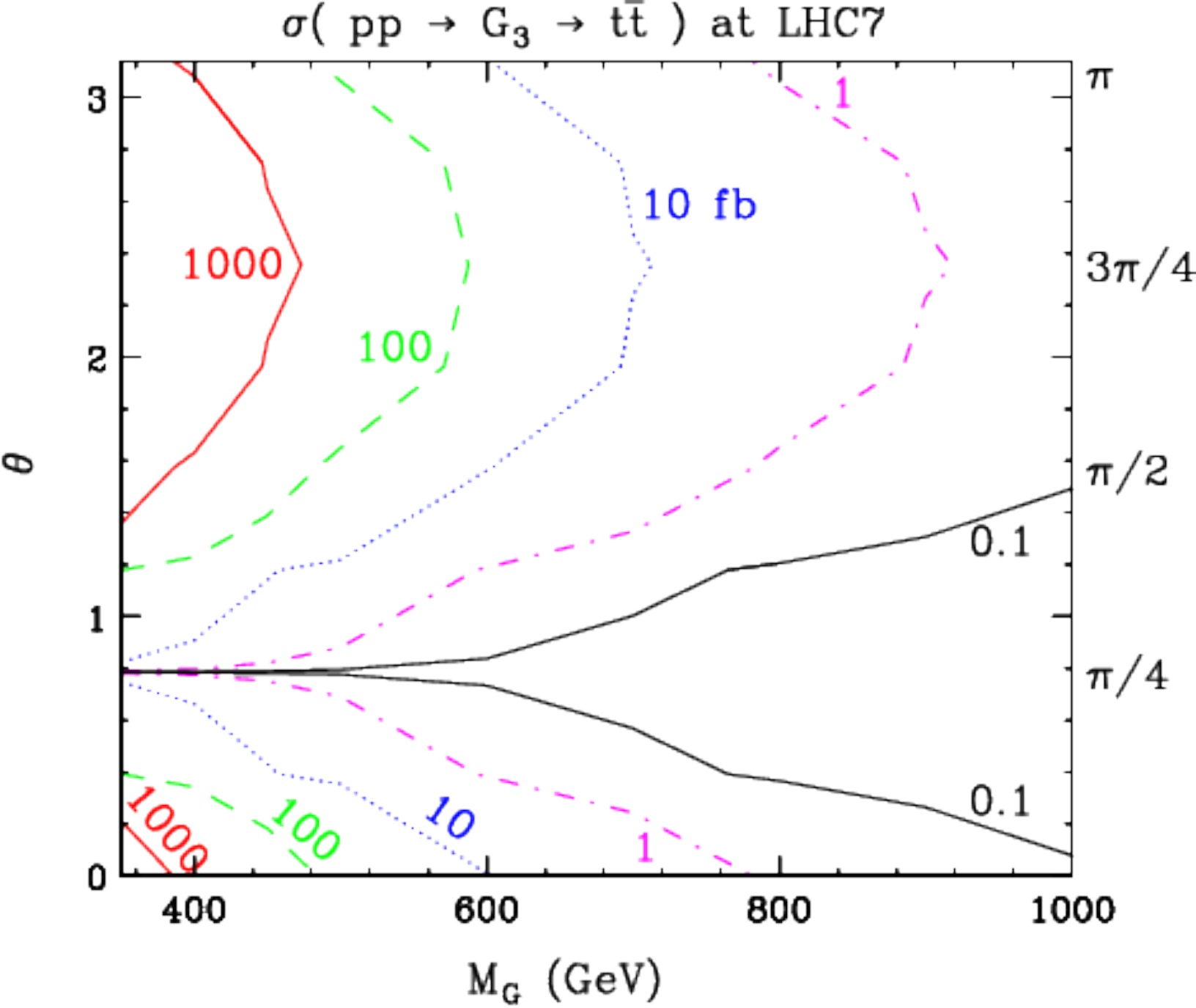}
    \caption{\label{sfig:G30j7tev}}
  \end{subfigure}
  \hskip1mm
  \begin{subfigure}[b]{0.49\textwidth}\centering
    \includegraphics[width=1\textwidth]{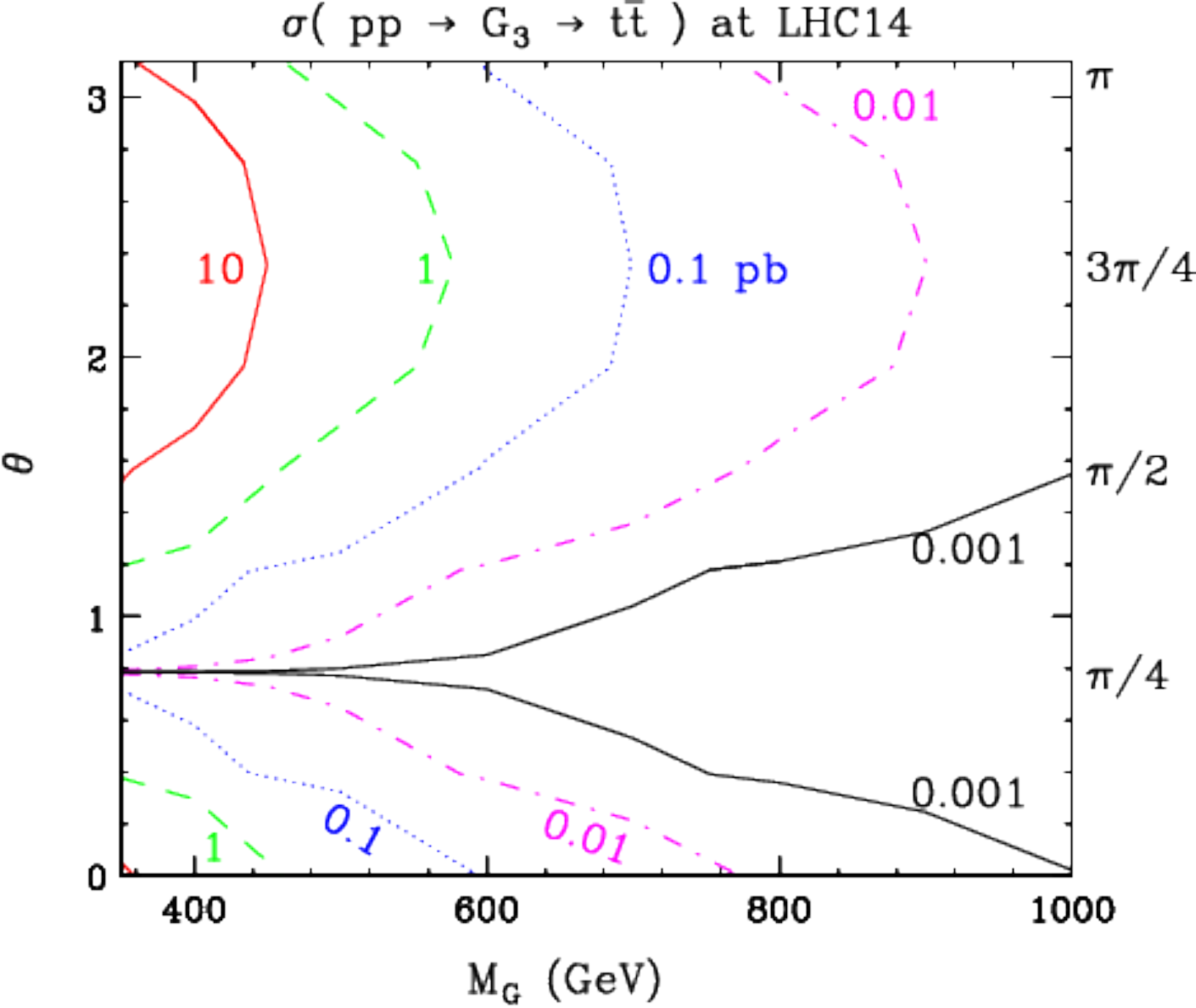}
    \caption{\label{sfig:G30j14tev}}
  \end{subfigure}
  \caption{\label{fig:G30j}
    Lines of constant production cross section in the $M_G$-$\theta$
    plane using $c_t=1$ for the loop induced (Born) processes
    $pp\to G_3\to t\bar t$ at proton--proton colliders of $7\tev$
    (\ref{sfig:G30j7tev}) and $14\tev$ (\ref{sfig:G30j14tev})
    center-of-mass energy.}
\end{figure}

\begin{figure}[t!]
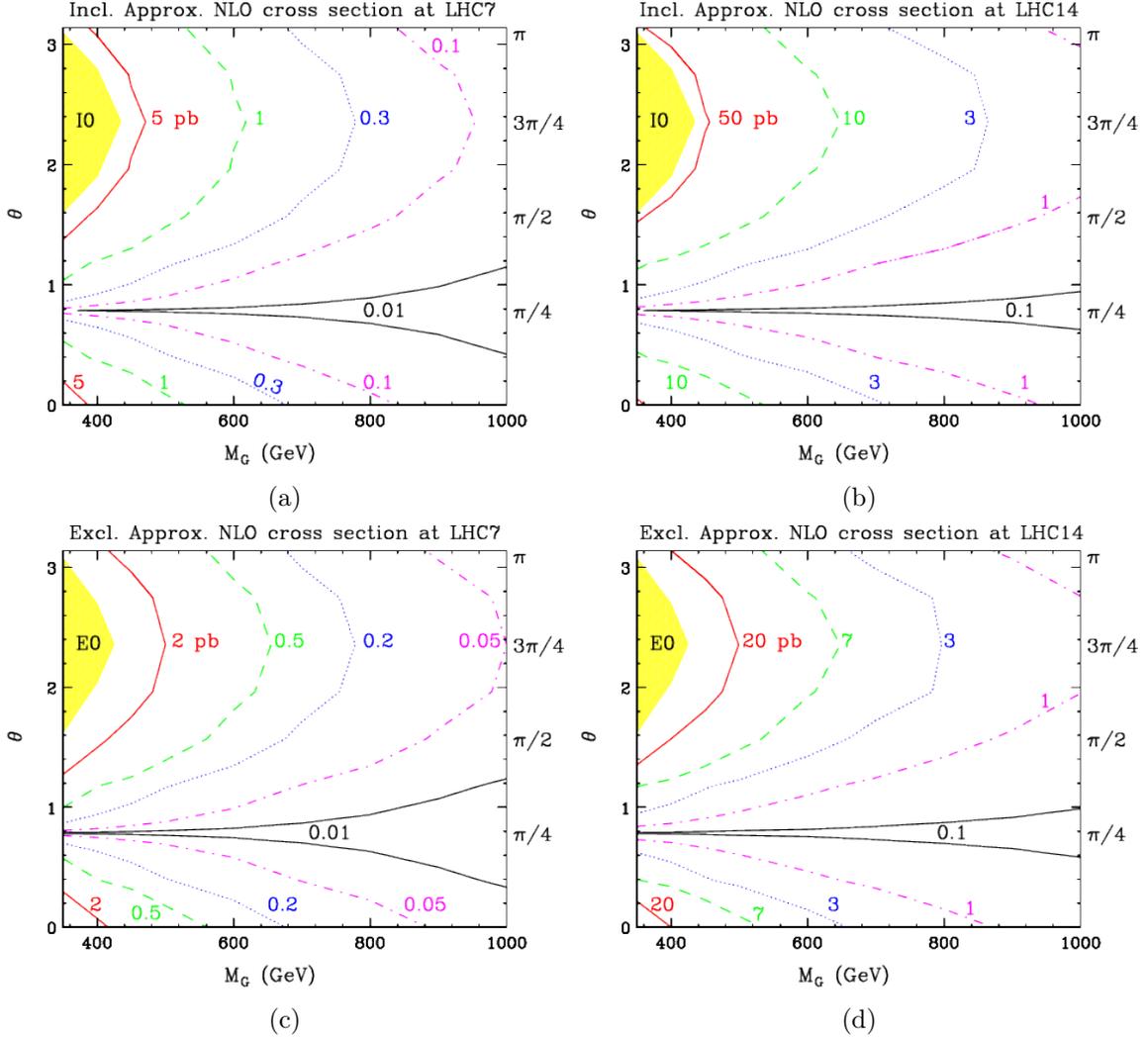

  \centering
  \begin{subfigure}[b]{0.49\textwidth}\centering
    \includegraphics[width=1\textwidth]{./FIGURES/%
      sigma_inc_approx_NLO_theta_MG3_7TeV.pdf}
    \caption{\label{sfig:G3inc7tev}}
  \end{subfigure}
  \hskip1mm
  \begin{subfigure}[b]{0.49\textwidth}\centering
    \includegraphics[width=1\textwidth]{./FIGURES/%
      sigma_inc_approx_NLO_theta_MG3_14TeV.pdf}
    \caption{\label{sfig:G3inc14tev}}
  \end{subfigure}
  \\[1mm]
  \begin{subfigure}[b]{0.49\textwidth}\centering
    \includegraphics[width=1\textwidth]{./FIGURES/%
      sigma_excl_approx_NLO_theta_MG3_7TeV.pdf}
    \caption{\label{sfig:G3exc7tev}}
  \end{subfigure}
  \hskip1mm
  \begin{subfigure}[b]{0.49\textwidth}\centering
    \includegraphics[width=1\textwidth]{./FIGURES/%
      sigma_excl_approx_NLO_theta_MG3_14TeV.pdf}
    \caption{\label{sfig:G3exc14tev}}
  \end{subfigure}
  \caption{\label{fig:G3approxNLO}
    Approximate NLO cross sections in the loop induced mode for the
    inclusive (upper row) and exclusive (lower row) production of
    $G_3\to t\bar t$ using $c_t=1$ at a $7\tev$~(\ref{sfig:G3inc7tev}
    and \ref{sfig:G3exc7tev}) and $14\tev$~(\ref{sfig:G3inc14tev} and
    \ref{sfig:G3exc14tev}) proton--proton collider. Regions of
    parameter space excluded at the one-sigma level are indicated by
    the colored areas, and labelled using the jet categories of
    Table~\ref{tab:limits}.}
\end{figure}

The bounds that apply to the $t\bar t+0\,j$ calculations are more
complicated to derive than in the previous case. As shown in
Eqs.~\eqref{eq:inc0limit} and \eqref{eq:exc0limit}, this is due to the
fact that there are non-negligible contributions from the $t\bar t+1\,j$
calculation that have to be taken into account once the extra jet is
not or does not get resolved.
Figure~\ref{fig:G30j} shows the results for the LO contribution of the
loop induced processes $pp\to G_3\to t\bar t$. To apply the
corresponding cross section limits, which we have worked out in
Table~\ref{tab:limits}, we however need the approximate NLO cross
section as explained in the previous section.\footnote{%
  The approximate NLO cross sections, i.e.~the left-hand sides of
  Eqs.~\eqref{eq:inc0limit} and~\eqref{eq:exc0limit}, shown in
  Figure~\ref{fig:G3approxNLO} have been evaluated neglecting the
  ${\cal O}(10\%)$ contribution from the $qg$ initial states in the
  calculation of the term $\sigma^\mrm{loose.cuts}_\mrm{Real}$, which
  is described by the $t\bar t+1$ parton process at LO and the
  application of loose jet cuts.}
These results are presented in Figure~\ref{fig:G3approxNLO}, including
the effect of the cross section bounds.
Figure~\ref{sfig:G3inc7tev} shows the inclusive case at $7\tev$ which is
described by Eq.~\eqref{eq:inc0limit}. The yellow shaded area denotes
the region in parameter space that is excluded by the inclusive
zero-jet measurement (I0) in Table~\ref{tab:limits}.
Figure~\ref{sfig:G3inc14tev} shows the theoretical cross section
predictions for $14\tev$. As before in Figure~\ref{fig:G31j}, only in
the $7\tev$ plot, the border of the excluded area can be associated
with a line of constant cross section. This does not apply to the
$14\tev$ plot.

Figures~\ref{sfig:G3exc7tev} and \ref{sfig:G3exc14tev} show the
exclusive case, where the theoretical predictions are calculated
according to Eq.~\eqref{eq:exc0limit} and the yellow shaded area
corresponds to the excluded area stemming from the exclusive zero-jet
measurement (E0) in Table~\ref{tab:limits}. The direct comparison
of Figure~\ref{fig:G30j} with Figures~\ref{sfig:G3exc7tev} and
\ref{sfig:G3exc14tev} reflects once more the important fact already
demonstrated in Figure~\ref{sfig:xsecG0jcompG1j}, namely that the
contribution of the one-jet case, where the jet is not resolved,
provides a substantial contribution to the zero-jet signal and cannot
be neglected in a realistic study. This is the reason why the zero-jet
cases provide constraints on the allowed parameter space that are
stronger than those of the one-jet cases. Nevertheless, in both jet bin
categories, the constraints on the parameter space are rather weak and
a large fraction of realistic scenarios cannot be excluded by present
LHC data. We will need more precise limits at $8\tev$ as well as
measurements at $14\tev$ in order to further constrain these types of
models.

\begin{figure}[t!]
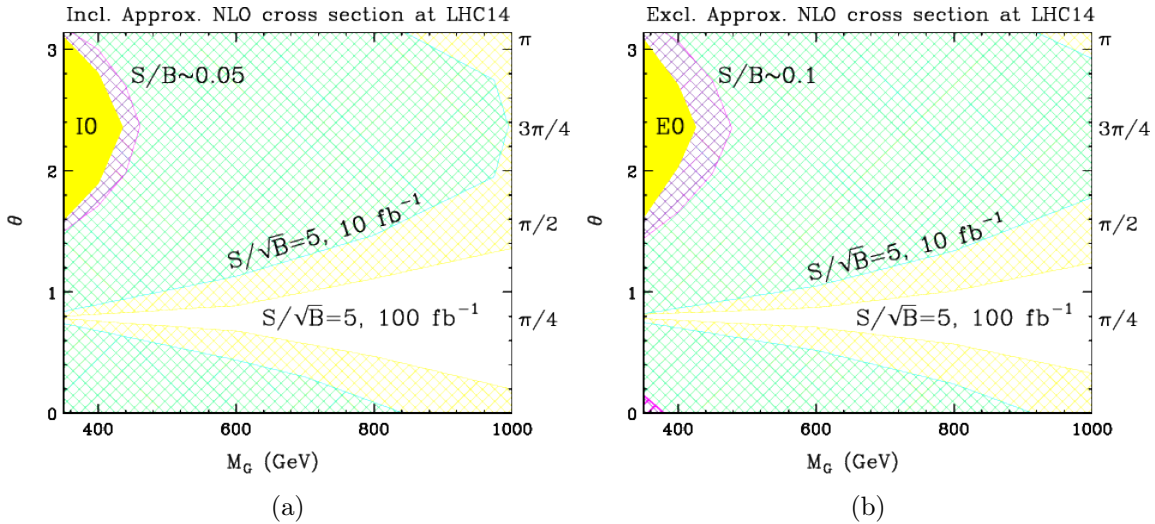

  \centering
  \begin{subfigure}[b]{0.49\textwidth}\centering
    \includegraphics[width=1\textwidth]{./FIGURES/%
      inc_approx_NLO_14TeV_projection.pdf}
    \caption{\label{sfig:I0prospects}}
  \end{subfigure}
  \hskip1mm
  \begin{subfigure}[b]{0.49\textwidth}\centering
    \includegraphics[width=1\textwidth]{./FIGURES/%
      excl_approx_NLO_14TeV_projection.pdf}
    \caption{\label{sfig:E0prospects}}
  \end{subfigure}
  \caption{\label{fig:prospects}
    Possible reach of a search for $G_3$ at the $14\tev$ LHC in the
    loop induced mode for the inclusive (\ref{sfig:I0prospects}) and
    exclusive (\ref{sfig:E0prospects}) production of $G_3\to t\bar t$
    using $c_t=1$. The shaded areas are used to indicate the one-sigma
    exclusions (in yellow) as in Figure~\ref{fig:G3approxNLO}, the
    constraints stemming from $S/B$ considerations (in purple) and the
    parameter space that is accessible with a statistical significance
    of five-sigma and beyond using $10\fbi$ (in green) and $100\fbi$
    (in light yellow) of LHC Run\:2 data.}
\end{figure}

\subsubsection{LHC Run\:2 prospects}

For the inclusive zero-jet case, the SM prediction for the top quark
pair production cross section is known up to NNLO in QCD (including
NNLL resummation effects)~\cite{Czakon:2013goa}, whereas for the
inclusive one-jet case, the cross section is computed at
NLO~\cite{Dittmaier:2007wz,Dittmaier:2008uj,Alioli:2011as}.
Based on these predictions given in the literature, we can estimate
the impact of the $G_3$ model on the $t\bar t+0,1$ jet final states.
We should however stress that we cannot expect to obtain more than
just a rough idea, since we build our arguments solely on a comparison
of rather inclusive cross sections, i.e.~no attempt is made to include
information from differential observables/properties that prove to be
different for the signal and the background. Comparing the SM
predictions, $\sigma_{t\bar t}=953.6\pb$ and $\sigma_{t\bar tj}=692\pb$,
with the results obtained in Figures~\ref{sfig:G3inc14tev} and
\ref{sfig:G31j14tev}, we find rather low $S/B$ ratios of maximally
$\sim5\%$ and $\sim1.4\%$ for the inclusive zero- and one-jet
category, respectively.\footnote{A somewhat better $S/B$ for the
  inclusive one-jet bin can be expected due to the differences in the
  jet parameters for the signal and background calculation.}
In the former case, the $S/B$ value corresponds to roughly the size of
the systematic uncertainty of the theoretical calculation, while in
the latter case it is considerably below the $6$-$9\%$ uncertainty of
the theory prediction.

Instead of considering an analysis that requires at least one jet, we
get a slightly better perspective by studying the situation in the
exclusive zero-jet bin. We obtain a theoretical prediction for this
contribution by subtracting the inclusive one-jet NLO result from the
NNLO prediction for $\sigma_{t\bar t}$. Using simple error
propagation, the uncertainty of this prediction amounts to
$11$-$14\%$. However, in a more sophisticated theoretical treatment, it
can be expected to control the prediction at the $10\%$ level or
below. This implies that the minimal $S/B$ ratio has to be of similar
size. The region in parameter space where this condition is fulfilled
is represented by the purple shaded area in Figure~\ref{sfig:E0prospects}.
The same requirement applied to the total inclusive cross section,
namely that $S/B\sim\Delta\sigma_{t\bar t}/\sigma_{t\bar t}\approx0.05$,
leads to weaker constraints on the parameters compared to the
exclusive case, cf.~Figure~\ref{sfig:I0prospects}.
The plots in Figure~\ref{fig:prospects} also show areas that
correspond to a statistical significance ($S/\sqrt B$) larger than
five, marking the discovery threshold. The areas shaded in green and
yellow correspond to integrated luminosities of $10\fbi$ and
$100\fbi$, respectively. We clearly see that in both the inclusive as
well as exclusive case, most of the parameter plane is already covered
after accumulating $10\fbi$ in Run\:2. This means that from a
statistical point of view the LHC has the ability to discover a $G_3$
resonance in a wide region of parameter space, even with a low
luminosity. We are not limited by a lack of statistics but by
relatively large systematic uncertainties. Therefore, a refined
strategy will have to exploit features in differential distributions
where we might be able to overcome the systematic uncertainties.

\section{Conclusions}\label{conclusion}

Searches for various types of resonances at the LHC are very well
motivated. These resonances may show up at an early stage of LHC
Run\:2. However the null results at Run\:1 indicate that their mass
scale may be high or they may be hiding in an exotic corner of the
phase space. The top quark is the heaviest particle in the Standard
Model. It decays promptly to a $W$ boson and a $b$-quark before
hadronization and may be vulnerable to new physics, which makes new
physics searches in the top quark sector very appealing.

In this paper, we investigated a top-philic resonance, called $G_3$,
which may be light. All existing analyses assume the production of a
$t\bar t$ resonance from $q\bar q$ annihilation, which depends on its
diquark coupling. We have proposed model-independent production
channels both at the tree level {\em and}\/ at the one-loop level. Our
studies show that in spite of being loop suppressed, the one-loop
channels dominate over the tree level processes and yield cross
sections of up to several picobarns, strongly dependent on the chosen
point in the model parameter space. While the tree level modes
generate more exotic final states featuring multi-top quark
signatures, the $G_3$ production via one loop leads to the appearance
of individual top quark pairs, often accompanied by jets. In
particular, the $t\bar t$ plus one-jet process is not suppressed
compared to the zero-jet process as one would naively expect by
counting orders in $\alpha_\mrm{s}$. It rather supersedes the zero-jet
case in its leading role and provides a non-negligible contribution to
both measurements in the one-jet bin and in the zero-jet bin,
depending on whether the additional jet is resolved or not. This is a
strong hint for the presence of large NLO corrections that affect the
$G_3\to t\bar t$ process and most likely lead to a significant
enhancement of the signal cross section. However, as a full NLO
treatment for this process is not feasible, we combined the $G_3$ and
$G_3\,j$ calculations in a specific way to generate an approximate NLO
prediction for the signal. The resulting jet cross sections should not
exceed the size of experimental uncertainties on recent top quark pair
measurements and hence were used to derive first constraints on the
parameters of the model. While we obtain more stringent constraints
from the inclusive zero-jet case, we yet find, regardless of the zero-
or one-jet condition, that the majority of the parameter space has not
been excluded by current LHC data. Certainly, stronger limits can be
achieved by a reduction of the experimental uncertainties in $t\bar t$
measurements. More likely, they may be obtained from the exploitation
of suitable differential distributions, although we want to emphasize
that for this study, we restricted ourselves to setting limits
according to the determination of total cross sections.

In our first attempt of obtaining estimates concerning the prospects
for LHC Run\:2, we utilized a simple $S/B$ and $S/\sqrt B$ statistics
method considering ordinary top quark production in pairs and in
association with jets as the only background sources for this
analysis. Studying the most promising channels (i.e.~the
$G_3\,t\bar t$ process yielding four top-quark final states and the
loop induced $G_3$ production yielding inclusive $t\bar t$ final
states), we identify the signal over background ratio, $S/B$, as the
constraining quantity in our analysis whereas the statistical
significance will be sufficiently large in all channels with just
$10$-$100\fbi$ of Run\:2 data. Figures~\ref{fig:G3_bound} and
\ref{fig:prospects} provide us with a sketch of the $14\tev$ LHC
sensitivity for each of these production channels.
We find that the $t\bar t$ cross sections provide very stringent
constraints on the axial coupling ($\theta=3\pi/4$) whereas the
multi-top quark production is particularly useful in testing the
parameter space away from the axial coupling regime. This is because
the $G_3\,t\bar t$ process does not have a chirality dependence.
We also observe that the $G_3\,t\bar t$ process is in much better
shape from an $S/B$ point of view than the loop level $t\bar t$
production, which suffers from larger backgrounds. However, the large
cross sections of the loop induced modes leave more space to improve
on $S/B$ by imposing suitable kinematical requirements. Moreover, the
reconstruction of the $G_3$ resonance will be easier than for the more
complicated final states of the multi-top quark production, and we
should stress that an alleviated reconstruction procedure can be very
useful in further background reduction. Taken the different strengths
of the production channels, we find them to be complementary in both
the $G_3$ discovery potential and the measurement of its properties.
One should therefore take advantage of each channel and revisit them
with full background analyses.

Finally, although we have considered a $t\bar t$ resonance, which
couples very weakly to the rest of the Standard Model, one could
anticipate a sizable interaction of the resonance to $b\bar b$ as
well. In this case, we end up with a richer spectrum of final states
including more diverse signatures such as $b\bar b\,b\bar b$,
$t\bar t\,b\bar b$, and $b\bar b+j$. Moreover, in the context of dark
matter searches, where we assume that the $G_3$ couples to the dark
matter in addition to its $t\bar t$ coupling, new channels open up
producing new signatures such as $t\bar t+\met$ and $j+\met$.
Including these new ideas, our first studies clearly show that a light
top-philic resonance is capable of producing a rich phenomenology in
final states at the LHC. We hence encourage the experimental
collaborations to pursue a dedicated study on it.

\acknowledgments 

NG wants to thank the other members of the \textsc{GoSam} collaboration
for their effort and for numerous valuable discussions.
KK is supported in part by the US DOE Grant DE-FG02-12ER41809 and by
the University of Kansas General Research Fund allocation 2301566.
SC is supported by the Basic Science Research Program through the
National Research Foundation of Korea funded by the Ministry of
Education, Science and Technology (2013R1A1A20 64120).


\bibliographystyle{JHEP}
\bibliography{refs}

\end{document}